\documentclass[10pt,twocolumn,twoside,journal]{IEEEtran}
\usepackage[usestackEOL]{stackengine}
\usepackage{cite}
\usepackage{amsmath}
\usepackage{amssymb}
\usepackage{amsfonts}
\usepackage{mathrsfs}
\usepackage{mathtools}
\usepackage[dvipsnames]{xcolor}
\usepackage[thmmarks, amsmath]{ntheorem}
\usepackage{enumitem}
\usepackage[bbgreekl]{mathbbol}
\usepackage{multirow}
\usepackage{booktabs}
\usepackage{float}
\usepackage{tikz}

\DeclareSymbolFontAlphabet{\mathbb}{AMSb}
\DeclareSymbolFontAlphabet{\mathbbl}{bbold}

\DeclareMathOperator{\diag}{diag}

\DeclareMathOperator{\diff}{d}

\DeclareMathOperator{\rank}{rank}

\DeclareMathOperator{\vspan}{span}

\providecommand{\norm}[1]{\lVert#1\rVert}
\newcommand{\R}{{\mathbb R}}

\newcommand{\mc}{\mathcal}
\newcommand{\ddt}{\tfrac{\diff}{\diff \!t}}
\newcommand{\ext}{\text{\upshape{ext}}}
\newcommand{\intn}{\text{\upshape{int}}}
\newcommand{\sync}{\text{\upshape{sync}}}

\newtheorem{property}{Property}
\newtheorem{theorem}{Theorem}
\newtheorem{lemma}{Lemma}

\newtheorem{assumption}{Assumption}

\newtheorem{definition}{Definition}

\newcommand\Blue[1]{{\color{blue}#1}}
\DeclareRobustCommand{\Wye}{\mathbin{
\tikz[x=1pt, y=1pt, scale=0.85]{\draw
    (0,0)    -- +(0, -4)      +(45:4) -- (0,0) -- +(135:4)
    ;}}}

\DeclareRobustCommand{\Wyegnd}{\mathbin{
\tikz[x=1pt, y=1pt, scale=0.85]{\draw
    (-1.4, 0) -- (1.4, 0)
    (-1, -1) -- (1, -1)
    (-0.4, -2) -- (0.4, -2)
    (0, 0) -- ++(0, 2) -- ++(-3, 0) coordinate (tmp)
    -- +(0, -4)
    (tmp) +(45:4) -- (tmp) -- +(135:4)
    ;}}}
\DeclareRobustCommand{\wye}{\mathbin{
\tikz[x=1pt, y=1pt, scale=0.7]{\draw
    (0,0)    -- +(0, -4)      +(45:4) -- (0,0) -- +(135:4)
    ;}}}

\DeclareRobustCommand{\wyegnd}{\mathbin{
\tikz[x=1pt, y=1pt, scale=0.7]{\draw
    (-1.4, 0) -- (1.4, 0)
    (-1, -1) -- (1, -1)
    (-0.4, -2) -- (0.4, -2)
    (0, 0) -- ++(0, 2) -- ++(-3, 0) coordinate (tmp)
    -- +(0, -4)
    (tmp) +(45:4) -- (tmp) -- +(135:4)
    ;}}}

\begin{document}
\title{
Grid-forming control of three-phase and single-phase converters across unbalanced transmission and distribution systems
}

\author{Shahin S. Nudehi, Dominic~Gro\ss{},~\IEEEmembership{Member,~IEEE}
\thanks{This material is based upon work supported by the National Science Foundation under Grant No. 2143188.}
\thanks{Shahin S. Nudehi and D. Gro\ss{} are with the Department of Electrical and Computer Engineering at the University of Wisconsin-Madison, Madison, USA; e-mail:\{nudehi,dominic.gross\}@wisc.edu }
}

\maketitle

\begin{abstract}
In this work, we investigate grid-forming control for power systems containing three-phase and single-phase converters connected to unbalanced distribution and transmission networks, investigate self-balancing between single-phase converters, and propose a novel balancing feedback for grid-forming control that explicitly allows to trade-off unbalances in voltage and power. We develop a quasi-steady-state power network model that allows to analyze the interactions between three-phase and single-phase power converters across transmission, distribution, and standard transformer interconnections. {We first investigate conditions under which this general network admits a well-posed kron-reduced quasi-steady-state network model. Our main contribution leverages this reduced-order model to develop analytical conditions for stability of the overall network with grid-forming three-phase and single-phase converters connected through standard transformer interconnections. Specifically, we provide conditions on the network topology under which (i) single-phase converters autonomously self-synchronize to a phase-balanced operating point and (ii) single-phase converters phase-balance through synchronization with three-phase converters. Moreover, we establish that the conditions can be relaxed if a phase-balancing feedback control is used.} Finally, case studies combining detailed models of transmission systems (i.e., IEEE 9-bus) and distribution systems (i.e., IEEE 13-bus) are used to illustrate the results for (i) a power system containing a mix of transmission and distribution connected converters and, (ii) a power system solely using distribution-connected converters at the grid edge.
\end{abstract}

\section{Introduction}

\IEEEPARstart {T}{he} rapid integration of renewable energy sources is resulting in an unprecedented period of change for the electric power system. At the heart of this transition to a zero-carbon power system is a paradigm shift from centralized fuel-based bulk generation using synchronous machines to renewable generation connected to transmission and distribution via power electronics. In this context, ensuring frequency stability of future converter-dominated power systems is a major concern. In contrast to synchronous machines that provide significant kinetic energy storage (i.e., inertia) and primary frequency control via the turbine governor system, renewables and power electronic converters deployed today do not contribute to frequency stabilization and jeopardize power system stability \cite{milano2018foundations,WEB+15,TvH16}.

Typically, power converters are broadly categorized into (i) \textit{grid-following} converters that assume a stable and slowly changing ac voltage frequency and magnitude at their point of connection and are controlled as current or power sources \cite{milano2018foundations}, and (ii) \textit{grid-forming} converters that impose a {well-defined} ac voltage {(i.e., magnitude and frequency)} at the point of connection \cite{MCC-DMD-RA:93,johnson2014synchronization,zhong2011synchronverters,line_dynamics}. Because of their stabilizing properties, grid-forming power converters are envisioned to replace synchronous machines as the cornerstone of future power systems \cite{milano2018foundations,LCP+2020,tayyebi2020frequency}.

The transition towards a power system largely operating based on renewable generation and power electronics is typically envisioned to follow two main pathways:
(i) replacing (centralized) conventional bulk generation with large-scale renewable plants such as wind farms and utility scale photovoltaics, and (ii) integrating distributed renewable energy resources into sub-transmission, distribution, and at the grid edge. While the first approach preserves today's system operating paradigms, the integration of significant renewable generation and controllable resources (i.e., single-phase and/or three-phase converters) in distribution systems and at the grid edge, results in significant challenges to standard operating paradigms, models, and analysis methods \cite{line_dynamics,tayyebi2020frequency,markovic2021understanding}.

Today's operating paradigms and analysis methods are based on the assumption that power flows from controllable bulk generation towards uncontrolled loads at the grid edge. Consequently, frequency stability is typically analyzed for balanced three-phase high-voltage systems with sub-transmission and distribution  modeled as a balanced aggregate load \cite{FV1988,tayyebi2020frequency,markovic2021understanding}. Conversely, voltage stability of distribution systems is typically analyzed under the assumption that the distribution substation can be modeled as slack bus (i.e., constant voltage magnitude and frequency) \cite{GR2020,wang2017existence,SHL2019,BZ2016}. Therefore, analytical results on frequency stability of multi-machine and/or converter-dominated systems typically either focus on balanced three-phase systems or single-phase microgrids. Although these approaches have proven themselves useful for today's power system, they may not provide sufficient insight for power systems that feature significant power generation and controllability at the grid edge. The exception are works that use simulations and experiments to establish that single-phase converters may self-balance when connected to a three-phase load \cite{Brian_Delta} and illustrate operation of single-phase converters in an islanded microgrid formed by a three-phase converter \cite{KOG2017}. Overall, to the best of our knowledge, no results are available in the literature that rigorously investigate stability of power system with grid-forming devices interacting across three-phase transmission, three-phase or single-phase distribution networks, and standard three-phase transformer interconnections.

A significant obstacle to developing stability conditions for power systems containing a mix of three-phase and single-phase converters is a lack of network models of suitable complexity for analytic stability studies. Notably, \cite{wang2017existence} provides conditions for the existence and uniqueness of power flow solutions for general multi-phase distribution systems with standard three-phase transformer interconnections but assumes that the substation is modeled as a slack bus (i.e., ideal voltage source) and does not develop models that are tractable for stability analysis. In \cite{bernstein2018load} several linear models for power analysis of three-phase distribution networks are developed but this work does not consider three-phase transformers or single-phase networks, and again assumes that the substation is modeled as a slack bus. Ultimately, to the best of our knowledge, the literature on power flow modeling does not establish models and methods that are tractable for analytic studies of the dynamic interactions between three-phase and single-phase power converters at different voltage levels and the impact of typical transformer interconnections on frequency synchronization, stability, and voltage unbalance.

Thus, to enable our small-signal stability analysis, this work first studies quasi-steady-state modeling of unbalanced power systems containing three-phase and single-phase networks. We develop a quasi-steady-state small-signal power flow model that is tractable for dynamic stability analysis and accounts for three-phase transmission, single-phase distribution, and common transformer interconnections. Notably, we provide conditions under which a well-posed reduced-order (i.e., Kron reduced  \cite{dorfler2012kron}) quasi-steady-state model exists and analytically characterize key properties needed for stability analysis. 

{Our main contribution leverages the reduced-order network model and its properties to develop analytical conditions for dynamic stability of frequency and voltage, and phase-balancing of three-phase and single-phase power converters deployed across transmission and distribution systems. In particular, our analytical stability conditions characterize the network topologies for which single-phase converters phase-balance through synchronization with three-phase converters. Moreover, we obtain analytical stability conditions that clarify the class of network topologies for which autonomous phase-balancing of single-phase converters occurs. This result explains the observations in \cite{Brian_Delta}  and generalizes them to general interconnections of three-phase and single-phase networks through standard transformer interconnections. Finally, we propose a phase-balancing feedback control that results in relaxed stability conditions and autonomously mitigate voltage unbalance at the primary control layer using local communication.}

{The prevalent method in the literature to voltage control and mitigating voltage unbalance in distribution systems is to leverage three-phase power flow optimization to adjust the reactive power injections of power converters modeled as controllable reactive power sources (i.e., grid-following control) \cite{Line,Rahul,Arnold, su2014optimal, karagiannopoulos2018}. This feedforward control approach generally requires accurate network models, i.e., may be sensitive to model errors, leads to large optimization problems solved by a centralized, i.e., may require extensive communication, and does not account for stability of the underlying controls. To reduce computation and communication requirements, feedback optimization for voltage control (but not phase-balancing) \cite{cavraro2022feedback} and  distributed feedback control methods for phase-balancing \cite{Bajo2015,Yao} have been proposed that assume a radial distribution network, balanced upstream network (i.e., the substation), and do not consider a mix of three-phase and single-phase networks, three-phase transformers, or dynamic stability of the converter controls. Finally, specialized costly devices such as three-phase static synchronous compensators \cite{Xu2010} or transformer online tap changers \cite{Liu2012} have been considered to for mitigating voltage unbalance but do not investigate their impact on dynamic grid stability. 

In contrast, the proposed phase-balancing feedback autonomously reduces voltage unbalance, does not require the network model, and uses local communication. The proposed control combines droop control \cite{MCC-DMD-RA:93, RLB+2012} for each phase with a phase-balancing feedback that can be tuned to adjust the power converter's contribution to mitigating voltage unbalance.}

In other words, the proposed control counteracts voltage unbalance using a combination of a primary control response and local communication. Moreover, the proposed control can conceptually be combined with the aforementioned centralized or distributed methods that infrequently update its setpoints. 

Finally, a detailed electromagnetic transient simulation of a power system consisting of a high-voltage transmission system and low-voltage distribution feeders is used to illustrate a system operation using a mix of transmission-connected three-phase converters and distribution-connected single-phase converters {and study voltage unbalances in a distribution feeder model for various scenarios.}

This manuscript is structured as follows. In Section~\ref{sec:problemsetup} we introduce and motivate the overall problem setup. Section~\ref{sec:networkmodel} develops a small-signal model of unbalanced multi-phase power systems including standard transformer topologies that are suitable for analytic stability studies. Section~\ref{Droop} discusses modeling of standard single-phase and three-phase grid-forming droop control in our modeling framework and also introduces the novel phase-balancing feedback. Analytic stability conditions for power systems that contain a mix of three-phase and single-phase grid-forming converters across transmission and distribution are introduced in Section~\ref{sec:stability}. Moreover, Section~\ref{sec:stability} introduces our generalized three-phase grid-forming droop control with phase-balancing control. Finally, section~\ref{Simulation} presents simulation studies and Section~\ref{sec:conclusion} provides the conclusions.

\subsection*{Mathematical notation}
We use $\mathbb{C}$ and $\R$ to denote the set of complex and real numbers. The complex imaginary unit is denoted by $j=\sqrt{-1}$, the element wise conjugate of a vector $y \in \mathbb{C}^n$ is denoted by $\overline{y}$, and $\operatorname{Re}(\mc Y)$ and $\operatorname{Im}(\mc Y)$ denote the real and imaginary part of a complex matrix $\mc Y$. Given a real matrix $A$, $A^\mathsf{T}$ denotes its transpose, and $A^{\dagger}$ denotes its pseudo-inverse. We write $A\succcurlyeq0$  $(A\succ0)$ to denote that $A$ is positive semidefinite (definite). For column vectors $x\in\R^n$ and $y\in\R^m$ we define $(x,y) = [x^\mathsf{T}, y^\mathsf{T}]^\mathsf{T} \in \R^{n+m}$, and $\norm{x}=\sqrt{x^\mathsf{T} x}$ denotes the Euclidean norm. Furthermore, $I_n$, $\mathbbl{0}_{n\times m}$, $\mathbbl{0}_{n}$, and $\mathbbl{1}_n$ denote the $n$-dimensional identity matrix, $n \times m$ zero matrix, and column vectors of zeros and ones of length $n$. The cardinality of a discrete set $\mc X$ is denoted by $|\mc X|$. Finally, $\odot$ and $\otimes$ denote the Hadamard product and Kronecker product.

{
\section{Motivation and Problem setup}\label{sec:problemsetup}
To introduce and motivate the problem setup considered in this work, this section briefly reviews modeling and control challenges that arise when integrating vast numbers of distributed renewable energy resources into sub-transmission, distribution, and at the grid edge.

\subsection{Modeling of transmission and distribution systems}
The transition towards a zero-carbon power system based on renewable generation results in the massive integration of distributed renewable energy resources and power electronics into sub-transmission, distribution, and at the grid edge. At the same time, today's power system models are based on the paradigm that power flows from transmission connected controllable bulk generation towards largely uncontrolled loads connected at lower voltage levels, e.g., in distribution systems. Based on this paradigm, transmission systems and distribution systems are typically modeled in isolation and using different models. In particular, transmission systems are typically modeled under the assumption that the system is balanced and sub-transmission and distributed are modeled as balanced aggregate loads \cite{FV1988,tayyebi2020frequency,markovic2021understanding}. At the same time, distribution systems are commonly modeled under the assumption that the distribution substation can be modeled as slack bus (i.e., constant voltage magnitude and frequency) \cite{GR2020,wang2017existence,SHL2019,BZ2016}.

As a consequence, stability analysis of transmission and distribution system focus on different aspects. While dynamic stability analysis of transmission systems often focuses on frequency stability, the frequency in distribution systems is often treated as constant and equal to the slack bus frequency and the focus is on studying steady-state voltage stability. 

Although, these approaches have proven themselves useful, the underlying models and analysis models are not suitable to study the dynamic interactions of grid-forming devices in networks that contain a mix of three-phase and single-phase networks across transmission and distribution (see Fig.~\ref{fig0}). Thus, the authors believe that modeling and analysis of the transmission and distribution grids can very likely not be separated in future systems. Moreover, recent results show that studying three-phase networks formed by grid-forming single-phase converters may be of interest in its own right \cite{Brian_Delta}. 

However, to the best of our knowledge no framework for modeling and analytic stability analysis exists that can capture the main salient features of interactions between grid-forming three-phase and single-phase devices across three-phase transmission and single-phase distribution.

\subsection{Control Objectives}
From the viewpoint of system control, a significant concern is how distributed renewable energy resources and power electronics integrated into distribution networks\Blue{,} and at the grid edge fit into the prevalent operating paradigms. In this work, we focus on grid-forming controls that provide standard primary control functions and enable power converters to contribute to the overall grid stability (i.e., both transmission and distribution grids). The control objectives are summarized below 
\begin{enumerate}
 \item During nominal steady-state operation the power converters need to  synchronize at the nominal frequency and a (balanced) operating point (i.e., ac voltage magnitude, and active and reactive power injection) periodically prescribed by a system-level controller.
 
 \item The converter should autonomously respond to variations in load or generation and contingencies to stabilize the grid ac voltage (i.e., magnitude and frequency) until a system-level controller can identify the new situation and provide an updated nominal steady-state operating point that is consistent with the post-contingency system configuration.
\end{enumerate}

To resolve the modeling challenges outlined in this section and develop a basis for control design and stability analysis, the next section develops a power network model that accounts for a mix of three-phase and single-phase networks and standard transformer interconnections. 

\begin{figure}[t]
\centering
\includegraphics[width=0.9\columnwidth]{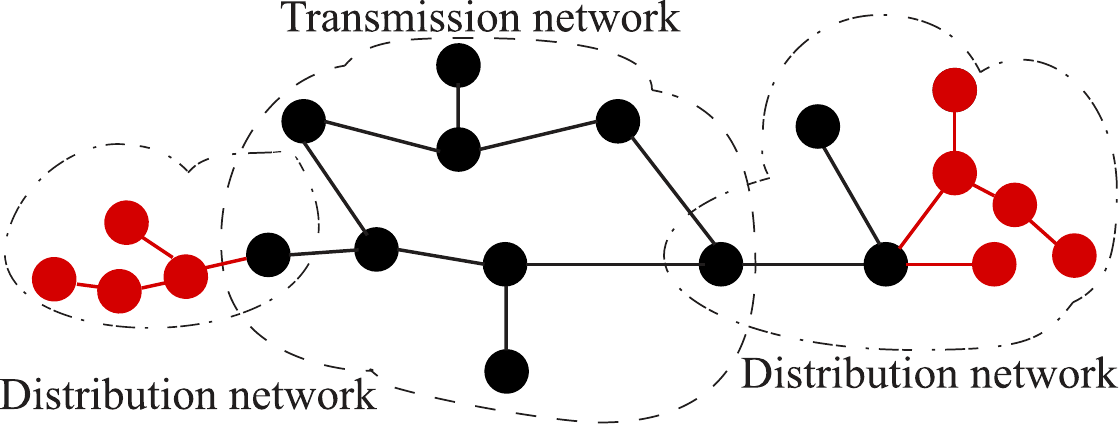}
\caption{Power network containing three-phase (black) and single-phase (red) buses and branches modeling transmission and distribution networks {and standard transformer interconnections (not shown)}.}
\label{fig0}
\end{figure}

}

\section{Power network model}\label{sec:networkmodel}
We consider a power system consisting of three-phase transmission and distribution including standard transformer interconnections, single-phase distribution, and single-phase and three-phase grid-forming power converters. This section introduces our basic assumptions on the network topology, the model of the power network and its branches, and analyzes basic properties of the power network model.

\subsection{Power network topology}\label{subsec:networktopology}
\begin{figure}[t]
\centering
\includegraphics[width=0.6\columnwidth]{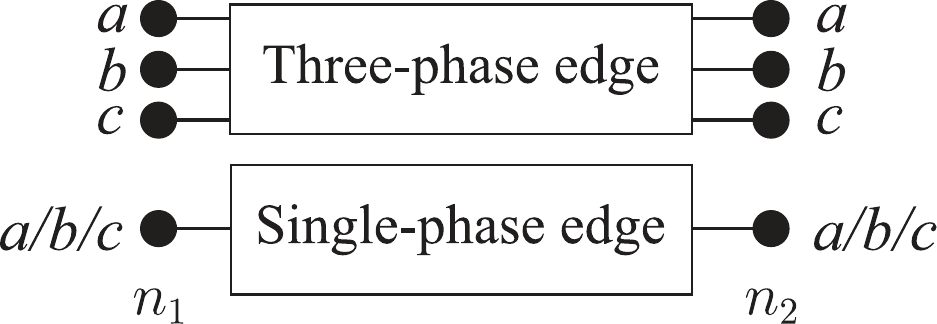}
\caption{Three-phase and single-phase edges connecting nodes $n_1$ and $n_2$.}
\label{fig1}
\end{figure}

{Standard graph-theoretic models used in stability analysis of power systems (see e.g., \cite{DSB2018}) commonly assume a balanced three-phase network or single-phase network. Therefore, to enable our stability analysis, we (i) extend standard graph theoretic models to model a mix of three-phase and single-phase networks or standard three-phase transformers, and (ii) rigorously define network topologies for which stability of a three-phase network with a mix of three-phase and single-phase converters can be guaranteed.

We assume} the network topology is modeled as a connected, directed graph $\mc G=(\mc N, \mc E)$. The nodes $\mc N \coloneqq \mc N_{1\phi} \cup \mc N_{3\phi}$ correspond to single-phase buses $\mc N_{1\phi}$ and three-phase buses $\mc N_{3\phi}$. The edges $\mc E \coloneqq \mc E_{3\phi}  \cup \mc E_{1\phi} \subseteq \mc N \times \mc N$ correspond to three-phase branches $\mc E_{3\phi}$ modeling lines and transformers and single-phase branches $\mc E_{1\phi}$ (see Fig.~\ref{fig0} and Fig.~\ref{fig1}). 

The nodes are partitioned into \emph{exterior} nodes $\mc N^{\ext}_{1\phi}$ and $\mc N^{\ext}_{3\phi}$ corresponding to buses with grid-forming converters connected on \emph{all phases}, and \emph{interior} nodes $\mc N^{\intn}_{1\phi}$ and $\mc N^{\intn}_{3\phi}$ that are not connected to grid-forming converters on \emph{any phase}. Three-phase edges $\mc E_{3\phi}  \subseteq \mc N_{3\phi} \times \mc N_{3\phi}$ connect three-phase nodes and the single-phase edges $\mc E_{1\phi} \subseteq \mc N_{1\phi} \times \mc N_{1\phi}  \cup  \mc N_{3\phi}$ connect a single-phase node $i \in \mc N_{1\phi}$ to a single-phase node $k \in \mc N_{1\phi}$ or one phase of a three-phase node $k \in \mc N_{3\phi}$. The three-phase edges $\mc E_{3\phi}$ are partitioned into branches $\mc E_{3\phi} \coloneqq \mc E_{\wye\wye} \cup \mc E_{\wyegnd\wyegnd} \cup \mc E_{\wyegnd\Delta} \cup \mc E_{\Delta\wye} \cup \mc E_{\Delta\Delta} \cup \mc E_{3\pi}$ corresponding to standard three-phase network branches (see Table~\ref{tab:1}).  The orientation of each edge $(i,k) \in \mc E$ specifies the orientation of the branch, e.g., $(i,k) \in \mc E_{\wyegnd\Delta}$ denotes that the primary terminal (i.e., $\Wyegnd$) of the branch is connected to node $i$ and the secondary terminal (i.e., $\Delta$) is connected to node $k$.

\begin{figure}[t!!]
\centering
\includegraphics[width=0.7\columnwidth]{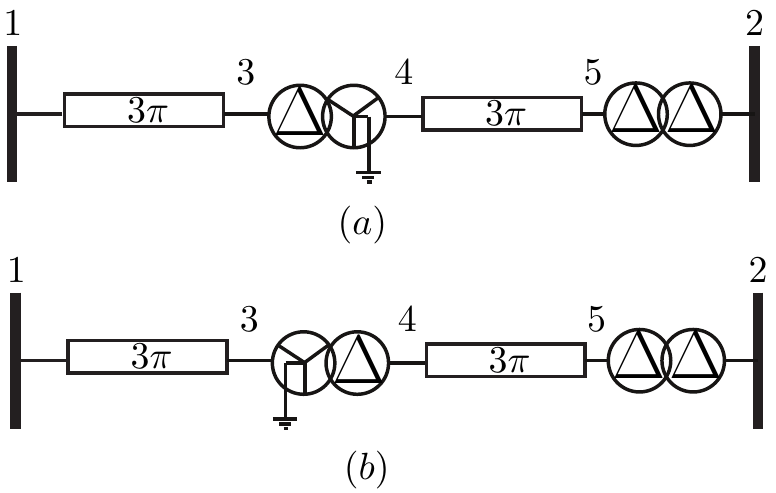}
\caption{The network ($a$) is {interior-exterior node connected} because, for every interior node (i.e., $3$, $4$, and $5$), there exists a path to the exterior node $1$ containing only three-phase lines and/or a $\Wyegnd\Delta$ transformers in the correct orientation. The network ($b$) is not {interior-exterior node connected} because the paths from the interior nodes $4$ and $5$ to the exterior nodes $1$ or $2$ either traverse a $\Delta\Delta$ transformer or a $\Wyegnd\Delta$ transformers from its secondary to its primary terminal.\label{fig:effective}}
\end{figure}

{To simplify the notation, we define the set $\mc E_\sync\coloneqq\mc E_{\wyegnd\wyegnd} \cup \mc E_{3\pi} \cup \mc E_{\wyegnd\Delta} \cup \mc E_{1\phi} \subseteq \mc E$ of network branches that play a crucial role in establishing stability of the overall power system. Broadly speaking, the power flowing across these network branches $\mc E_\sync$ sufficiently captures angle and voltage differences between its terminals to ensure synchronization of converters through the network. In contrast, the power flowing across the remaining branches $\mc E \setminus \mc E_\sync$ only partially reflects angle and voltage differences between their terminals. Moreover, the following definition is crucial to our analysis.}
\begin{definition} {\bf({interior-exterior node connected} network)} \label{def:1}
The network $\mc G$ is {interior-exterior node connected} if, for any interior node $k \in \mc N^{\intn}$, the subgraph $\mc G_\sync\coloneqq(\mc N, \mc E_\sync) \subseteq \mc G$ contains a path to an exterior node $l \in \mc N^{\ext}_{3\phi}$ that, starting from $k \in \mc N^{\intn}$, traverses all edges from their primary terminal to their secondary terminal.
\end{definition}
{An illustrative example is shown in Fig.~\ref{fig:effective}. Broadly speaking, interior-exterior node connectivity in the sense of Definition~\ref{def:1} ensures that interior node voltages can be recovered from exterior node voltages and bus power injections. 

We emphasize that interior-exterior node connectivity does not imply connectivity of the overall graph $\mc G$. Moreover, interior-exterior node connectivity is related to the notion of an effectively grounded three-phase system encountered in the literature on protection systems that ensures a low impedance path for fault current \cite{gonen2008electric}. Specifically, if the network associated with the graph  $\mc G_\sync$ is effectively grounded, then it is interior-exterior node connected. However the converse is not true and the network associated with the graph $\mc G$ may be effectively grounded through branches not contained in $\mc G_\sync$. Nonetheless, this highlights that assuming the graph to be interior-exterior node connected is not overly restrictive. Similar assumptions have been made in the literature to, e.g., ensure the existence and uniqueness of power flow solutions in multi-phase distribution systems that only contain constant power sources and model the substation transformer as a slack bus \cite[Sec. IV-C]{wang2017existence}. Our definition can be understood as a generalization to more general networks.}

Next, we restrict the power network topology to ensure that the power network models developed in the remainder of this manuscript are well-posed.
\begin{assumption}{\bf(Well-posed network)} We assume that \label{assum:wp}
\begin{enumerate}[label=\arabic*., ref=\theenumi]
 \item $\mc G$ is simple, connected, and {interior-exterior node connected}, \label{assum:wp:1}
 \item the number of transformer branches of a specific type (i.e., $\mc E_{\wye\wye}$, $\mc E_{\wyegnd\wyegnd}$, $\mc E_{\wyegnd\Delta}$, $\mc E_{\Delta\wye}$, and $\mc E_{\Delta\Delta}$), voltage ratio, and orientation traversed by any path between any two nodes $(k,l) \in \mc N_{3\phi} \times \mc N_{3\phi}$ is identical, \label{assum:wp:2}
 \item the graph $\mc G_{1\phi}\coloneqq(\mc N,\mc E_{1\phi})$ contains no path connecting any three-phase nodes $(i,k) \in \mc N_{3\phi} \times \mc N_{3\phi}$. \label{assum:wp:3}
 \end{enumerate}
\end{assumption}

\begin{figure}[t]
\centering
\includegraphics[width=0.6\columnwidth]{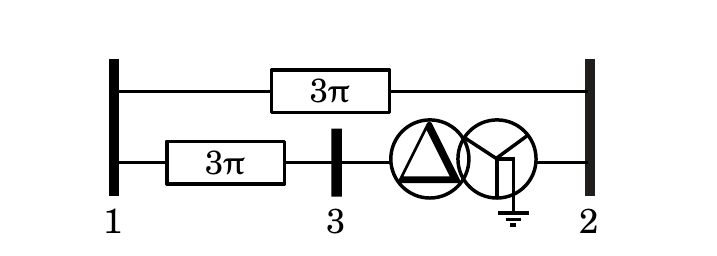}
\caption{Network with inconsistent phase shift between the upper path ($0^\circ$ phase shift) and lower path ($30^{\circ}$ phase shift) between bus $1$ and bus $2$. \label{fig:phase_incons}}
\end{figure}

Assumption~\ref{assum:wp}.\ref{assum:wp:1} requires that the network is {interior-exterior node connected}, connected, and there is at most one edge between any two nodes\footnote{Considering Assumption~\ref{assum:wp}.\ref{assum:wp:2}, parallel edges of the same type (e.g., double lines) can be combined into one edge.}. Assumption~\ref{assum:wp}.\ref{assum:wp:2} and Assumption~\ref{assum:wp}.\ref{assum:wp:3} rule out connections with inconsistent phase shifts, voltage ratios, inconsistent phase configuration, and connections between three-phase nodes through single-phase networks (see Fig.~\ref{fig:phase_incons} for a network with inconsistent phase shift). Finally, for each $(i,k) \in  \mc N_{1\phi} \times \mc N_{3\phi}$, we define the incidence vector $\mc I_{i,k} \in \{0,1\}^3$ with one non-zero entry that models its phase configuration (e.g., $\mc I_{i,k}=(1,0,0)$ if it is connected to phase $a$ of $k \in \mc N_{3\phi}$). For each $(i,k) \in  \mc N_{1\phi} \times \mc N_{1\phi}$ we define $\mc I_{i,k}\coloneqq 1$.

\subsection{Quasi-steady-state branch model}\label{subsec:branchadmittance}
{To set up our notation and clarify our basic assumptions, this section briefly reviews quasi-steady-state models of typical network branches (see \cite{chen1991three} for further details).} 
To each node $i \in \mc N$, we associate a voltage phasor $V_i=v_i e^{\theta_i} \in \mathbb{C}^{n_i}$, where $n_i=3$ if $i \in \mc N_{3\phi}$ and $n_i=1$ if $i \in \mc N_{1\phi}$. To each branch $(i,k) \in \mc E$, we associate the currents $I_{i,k} \in \mathbb{C}^{n_i}$ and $I_{k,i} \in \mathbb{C}^{n_k}$ flowing into the branch primary terminal and secondary terminal, respectively. The branch currents are given by the branch admittance matrix $Y_{i,k} \in \mathbb{C}^{n_i+n_k \times n_i+n_k}$, and
\begin{align}\label{eq:branchaddmittancemodel}
\begin{bmatrix} I_{i,k} \\ I_{k,i} \end{bmatrix} = \underbrace{j b_{i,k}\begin{bmatrix}
	\mathsf{Y}_{\mathsf{ii}} & \mathsf{Y}_{\mathsf{ik}}\\
	\mathsf{Y}_{\mathsf{ik}}^\mathsf{T} & \mathsf{Y}_{\mathsf{kk}}
\end{bmatrix}}_{\eqqcolon Y_{i,k}} \begin{bmatrix} V_i \\ V_k \end{bmatrix}.
\end{align}
The entries $\mathsf{Y}_{\mathsf{ii}} \in \mathbb{C}^{n_i \times n_i}$, $\mathsf{Y}_{\mathsf{ik}} \in \mathbb{C}^{n_i \times n_k}$, $\mathsf{Y}_{\mathsf{kk}} \in \mathbb{C}^{n_k \times n_k}$ of the admittance matrix $Y_{i,k}$ for common network branches can be obtained using the approach in \cite{chen1991three} and are given by Table~\ref{tab:1},
\begin{table}[bt!!]\centering
\caption{Admittance matrices of common network branches}
\begin{tabular}{|l|c|c|c|c|} \hline
Type & Edge set & $\mathsf{Y}_\mathsf{ii}$ & $\mathsf{Y}_\mathsf{kk}$ &$\mathsf{Y}_\mathsf{ik}$ \\
\hline
 $\Wyegnd\Wyegnd$ transformer & $\mc E_{\wyegnd\wyegnd}$ & $I_3$ & $I_3$ & $-I_3$ \\
\hline
 $\Wyegnd\Wye$ transformer & $\mc E_{\wyegnd\wye}$ & $\mathsf{Y_1}$ & $\mathsf{Y_1}$ & $-\mathsf{Y_1}$ \\
\hline
 $\Wyegnd\Delta$ transformer & $\mc E_{\wyegnd\Delta}$ & $I_3$ & $\mathsf{Y_1}$ & $\mathsf{Y_2}$ \\
\hline
$\Wye\Wye$ transformer & $\mc E_{\wye\wye}$ & $\mathsf{Y_1}$ & $\mathsf{Y_1}$ & $-\mathsf{Y_1}$ \\
\hline
 $\Wye\Delta$ transformer & $\mc E_{\wye\Delta}$ & $\mathsf{Y_1}$ & $\mathsf{Y_1}$ & $\mathsf{Y_2}$ \\
\hline
$\Delta\Delta$ transformer & $\mc E_{\Delta\Delta}$ & $\mathsf{Y_1}$ & $\mathsf{Y_1}$ & $-\mathsf{Y_1}$ \\
\hline
Three-phase line & $\mc E_{3\pi}$ & $I_3$ & $I_3$ & $-I_3$\\
\hline
Single-phase branch & $\mc E_{1\phi}$ & $1$ & $\mc I_{i,k}^\mathsf{T} \mc I_{i,k}^{\phantom{\mathsf{T}}}$ & $-\mc I_{i,k}^\mathsf{T} \mc I_{i,k}^{\phantom{\mathsf{T}}}$\\
\hline
\end{tabular}
\label{tab:1}
\end{table}
the susceptances $b_{i,k} \in \mathbb{R}_{>0}$, and the matrices
\begin{align*}
\mathsf{Y_1}\coloneqq\frac{1}{3}\begin{bmatrix}
2 & -1 & -1\\
-1 & 2 & -1\\
-1 & -1 & 2
\end{bmatrix},\quad \mathsf{Y_2}\coloneqq\frac{1}{3}\begin{bmatrix}
-1 & 1 & 0\\
0 & -1 & 1\\
1 & 0 & -1
\end{bmatrix}.
\end{align*}

{The branch flow model \eqref{eq:branchaddmittancemodel} assumes that all branches are lossless. This assumption is standard in analytical studies of power system dynamics and typically justified for transmission lines and three-phase converters. However, it is debatable for low-voltage distribution lines that typically have non-negligible resistance. Accounting for, or compensating, the impact of distribution lines with non-negligible resistance is seen as interesting topic for future work. In particular, in converter-dominated power system the effect of line resistance can, conceptually, be compensated through the converter \cite{RLB+2012,line_dynamics}.}

Finally, the power $S_{i,k} \in \mathbb{C}^{n_i}$ flowing into of the branch primary terminal and the power $S_{k,i} \in \mathbb{C}^{n_k}$ flowing into its secondary terminal are given by
\begin{align}
S_{i,k} =  {\tfrac{1}{2}} V_i \odot \overline{I}_{i,k}, \quad S_{k,i} = {\tfrac{1}{2}} V_k \odot \overline{I}_{k,i}. \label{eq:pf}
\end{align}

\subsection{Linearized branch power flows} \label{subsec:branchpf}
{In this section, we develop linearized branch power flow models for various standard elements and preliminary analytic results.} To obtain a network model that is tractable for analytical stability studies, we linearize the real and imaginary part of the complex powers \eqref{eq:pf} at the nominal voltage magnitude $v^\star_i \in \mathbb{R}^{n_i}$ (i.e., $1~\mathrm{pu}$ for each connected phase) and {a} phase angle configuration $\theta^\star_i \in \mathbb{R}^{n_i}$ and $\theta^\star_k \in \mathbb{R}^{n_k}$ corresponding to zero power flow across each branch\footnote{Existence of bus voltages for the entire network with nominal magnitude and zero power flow across each branch is implied by Assumption~\ref{assum:wp}.}. This approach is justified within standard voltage limits and for small phase angle differences across edges $\mc E$ and results in
\begin{align*}
 \begin{bmatrix}
  S_{\delta,i,k}\\
  S_{\delta,k,i}
 \end{bmatrix}=\underbrace{ b_{i,k} \begin{bmatrix}
	\mathsf{J}_{\mathsf{ii}} \\ \mathsf{J}_{\mathsf{ik}} \end{bmatrix} \begin{bmatrix}
	\mathsf{J}_{\mathsf{ii}}^\mathsf{T} & \mathsf{J}_{\mathsf{ik}}^\mathsf{T}\end{bmatrix}}_{\eqqcolon J_{i,k}}
 \begin{bmatrix}
    V_{\delta,i}\\
    V_{\delta,k}
 \end{bmatrix},
\end{align*}
with branch power flow matrix $J_{i,k} \in \mathbb{R}^{2(n_i+n_k) \times 2(n_i+n_k)}$, real and reactive powers $S_{\delta,i,k}=(P_{\delta,i,k},Q_{\delta,i,k})  \in \mathbb{R}^{2 n_i}$, $S_{\delta,k,i}=(P_{\delta,k,i},Q_{\delta,k,i})  \in \mathbb{R}^{2 n_k}$, and voltage phase angles and magnitudes $V_{\delta,i}=(\theta_{\delta,i},v_{\delta,i}) \in \mathbb{R}^{2 n_i}$. Moreover, the matrices $\mathsf{J}_{\mathsf{ii}} \in \mathbb{R}^{2 n_i \times 2 n_i}$ and $\mathsf{J}_{\mathsf{ik}}\in \mathbb{R}^{2 n_i \times 2 n_k}$ are further partitioned into
\begin{align}\label{eq:9b}
\mathsf{J}_{\mathsf{ii}}=\begin{bmatrix}	\mathsf{P}_{\mathsf{ii}}& -\mathsf{R}^\mathsf{T}_{\mathsf{ii}}\\
	-\mathsf{R}_{\mathsf{ii}}& \mathsf{P}_{\mathsf{ii}}
\end{bmatrix}, \quad \mathsf{J}_{\mathsf{ik}}=\begin{bmatrix}
	-\mathsf{P}_{\mathsf{ik}}& -\mathsf{R}_{\mathsf{ik}}\\[0.2em]
	\mathsf{R}_{\mathsf{ik}}& -\mathsf{P}_{\mathsf{ik}}
\end{bmatrix}
\end{align}
\begin{table}[tb!!!]\centering
\caption{Linearized power flow model of standard network branches}
\begin{tabular}{|c|c|c|c|c|} \hline
Type & $\mathsf{P}_\mathsf{ii}$ & $\mathsf{R}_\mathsf{ii}$ &$\mathsf{P}_\mathsf{ik}$ & $\mathsf{R}_\mathsf{ik}$ \\
\hline
 $\Wyegnd\Wyegnd$ transformer & $I_3$ & $\mathbbl{0}_{3\times 3}$ & $I_3$ & $\mathbbl{0}_{3\times 3}$\\
\hline
 $\Wyegnd\Wye$ transformer & $\mathsf{P_1}$ & $\mathsf{P_2}$ & $\mathsf{P_1}$ & $\mathsf{P_2}$\\
\hline
 $\Wyegnd\Delta$ transformer & $I_3$ & $\mathbbl{0}_{3\times 3}$ & $\mathsf{P_3}$ & $\mathsf{P_4}$\\
\hline
 $\Wye\Wye$ transformer & $\mathsf{P_1}$ & $\mathsf{P_2}$ & $\mathsf{P_1}$ & $\mathsf{P_2}$\\
\hline
$\Wye\Delta$ transformer & $\mathsf{P_1}$ & $\mathsf{P_2}$ & $\mathsf{P_3}$ & $\mathsf{P_4}$\\
\hline
$\Delta\Delta$ transformer & $\mathsf{P_1}$ & $\mathsf{P_2}$ & $\mathsf{P_1}$ & $\mathsf{P_2}$\\
\hline
Three-phase line & $I_3$ & $\mathbbl{0}_{3\times 3}$ & $I_3$ & $\mathbbl{0}_{3\times 3}$\\
\hline
Single-phase branch & $1$ & $0$ & $\mc I_{i,k}$ & $\mathbbl{0}_{n_k \times 1}$ \\
\hline
\end{tabular}
\label{tab:2}
\end{table}
and defined by Table~\ref{tab:2} and the matrices
\begin{subequations}\label{eq:pfmatrices}
\begin{align}
\mathsf{P_1}\coloneqq\frac{1}{{12}}\begin{bmatrix}
4 & 1  & 1\\
1 & 4 & 1\\
1 & 1 & 4
\end{bmatrix}, \quad
\mathsf{P_2}\coloneqq\frac{\sqrt3}{{12}}\begin{bmatrix}
0 & 1 & -1\\
-1 & 0 & 1\\
1 & -1 & 0
\end{bmatrix},\\
\mathsf{P_3}\coloneqq\frac{1}{{4}}\begin{bmatrix}
1 & 0 & 1\\
1 & 1 & 0\\
0 & 1 & 1
\end{bmatrix}, \quad
\mathsf{P_4}\coloneqq\frac{\sqrt3}{{12}}\begin{bmatrix}
1 & 0 & -1\\
-1 & 1 & 0\\
0 & -1 & 1
\end{bmatrix}.
\end{align}
\end{subequations}

Letting $\nu_n=(\mathbbl{1}_n,\mathbbl{0}_n)$ the following Proposition directly follows from Table~\ref{tab:2} and  \eqref{eq:pfmatrices}.
\begin{property}{\bf{(Branch power flow matrices)}}\label{property:branchproperties}
For all $(i,k)\in\mc E$ it holds that $J_{i,k}\succcurlyeq0$, $\mathsf{J}_\mathsf{ii}\succcurlyeq0$, and $J_{i,k}(\nu_{n_i},\nu_{n_k})=\mathbbl{0}_{2(n_i + n_k)}$. Moreover, for all $(i,k) \in \mc E_\sync$ it holds that $\mathsf{J}_{\mathsf{ii}} \succ 0$ and, by construction, $\mathsf{J}_{\mathsf{ik}}\mathsf{J}_{\mathsf{ik}}^\mathsf{T}-\mathsf{J}_{\mathsf{ik}} \mathsf{J}_{\mathsf{ii}} (\mathsf{J}_{\mathsf{ii}} \mathsf{J}_{\mathsf{ii}}^\mathsf{T})^{-1} \mathsf{J}_{\mathsf{ii}} \mathsf{J}_{\mathsf{ik}}^\mathsf{T}= \mathbbl{0}_{2 n_k \times 2 n_k}$, i.e., the Schur complement of $J_{i,k}$ with respect to the block $b_{i,j} \mathsf{J}_{\mathsf{ii}} \mathsf{J}_{\mathsf{ii}}^\mathsf{T}$ is a zero matrix.
\end{property}

\subsection{Linearized power network model}\label{subsec:networkpf}
{This section extends graph-theoretic modeling and analysis techniques to formally establish properties of power networks that contain three-phase networks, single-phase networks, and standard transformer interconnections. These preliminary results form the basis for developing analytical stability certificates in Section~\ref{sec:stability}.} The active and reactive power injection at every bus $i \in \mc N$ is given by $S_{\delta,i} \coloneqq \sum_{k: (i,k) \in \mc E} S_{\delta,i,k} \in \mathbb{R}^{2 n_i}$. The overall vector of bus power injections and voltage deviations is denoted by $S_\delta \coloneqq (S_{\delta,1},\ldots,S_{\delta,n}) \in \mathbb{R}^{n_{\mc N}}$ and $V_\delta \coloneqq (V_{\delta,1},\ldots,V_{\delta,n}) \in  \mathbb{R}^{n_{\mc N}}$, and $n_{\mc N} \coloneqq 6 |\mc N_{3\phi}|+ 2 |\mc N_{1\phi}|$ denotes the dimension of the nodal voltages / powers. Moreover, $n_{\mc E} \coloneqq 6 |\mc E_{3\phi}|+ 2 |\mc E_{1\phi}|$ denotes the total number of branch phases. To model the overall network, the \emph{generalized oriented incidence matrix} $B \in \mathbb{R}^{n_{\mc N} \times n_{\mc E}}$ is defined as
\begin{align}
 B \coloneqq \begin{bmatrix} B_1 & \ldots & B_{|\mc E|} \end{bmatrix} = \begin{bmatrix} B_{1,1} & \ldots & B_{1,|\mc E|} \\
      \vdots & \ddots & \vdots \\
      B_{|\mc N|,1} & \ldots & B_{|\mc N|,|\mc E|}
     \end{bmatrix},
\end{align}
i.e., every column of $B$ corresponds to an edge in the graph $\mc G$ and models the location and type of a branch in the network. To this end, we use $(l_i,l_k) \in \mc E$ to denote the indices of the nodes connected by the edge with index $l \in \mathbb{N}_{[1,|\mc E|]}$ and $n_l$ to denote the number of phases (i.e., $n_l=1$ if $(l_i,l_k) \in \mc E_{1\phi}$ and $n_l=3$ if $(l_i,l_k) \in \mc E_{3\phi}$). Each matrix $B_l \in \mathbb{R}^{n_{\mc N} \times 2 n_l}$ corresponds to an edge $l \in \mathbb{N}_{[1,|\mc E|]}$ in the graph and can be further partitioned into blocks $B_{i,l} \in \mathbb{R}^{2n_i \times 2n_l}$ that correspond to the node connections, i.e., $B_{l_i,l}=\mathsf{J}_\mathsf{l_i l_i}$ and $B_{l_k,l}=\mathsf{J}_\mathsf{l_i l_k}$ and $B_{j,l}=\mathbbl{0}_{2n_i \times 2n_l}$ in all other cases. The overall power network model is given by the susceptance matrix $W=\diag\{b_{l_i,l_k} I_{2 n_l}\}_{l=1}^{|\mc E|} \succ 0$ and
\begin{align}\label{eq:networkpf}
 S_\delta = \underbrace{B W B^\mathsf{T}}_{\eqqcolon J} V_\delta = \sum\nolimits_{l=1}^{|\mc E|} \underbrace{b_{l_i,l_k} B_l B_l^\mathsf{T}}_{\eqqcolon \mc J_l} V_\delta.
\end{align}
Moreover, we again partition the nodes into \emph{external} and \emph{internal} nodes and partition the matrix $J$ accordingly to obtain
\begin{align}\label{eq:networkpowerflow}
 \underbrace{\begin{bmatrix}
  S^{\ext}_{\delta}\\
  S^{\intn}_{\delta}
   \end{bmatrix}}_{= S_{\delta}}&=
   \underbrace{\begin{bmatrix}
    J_{\ext} & J_{\text{\upshape{c}}} \\ J^\mathsf{T}_{\text{\upshape{c}}} & J_{\intn}
   \end{bmatrix}}_{= J}
   \underbrace{\begin{bmatrix}
V^{\ext}_{\delta}\\
V^{\intn}_{\delta}
\end{bmatrix}}_{= V_{\delta}},
\end{align}
where $S^{\ext}_\delta \in \mathbb{R}^{n_{\mc N^\ext}}$, $n_{\mc N^\ext} \coloneqq 6 |\mc N^\ext_{3\phi}|+ 2 |\mc N^\ext_{1\phi}|$ and $S^{\intn}_\delta \in \mathbb{R}^{n_{\mc N^\intn}}$, $n_{\mc N^\intn}=n_{\mc N} - n_{\mc N^\ext}$ are the power injections at the interior and exterior buses.
It directly follows from \eqref{eq:networkpf} and Property \ref{property:branchproperties} that $\mc J_l \succcurlyeq 0$, $J \succcurlyeq 0$, and $J (\nu_{n_1},\ldots,\nu_{n_{|\mc N|}})=\mathbbl{0}_{n_{\mc N}}$.
For brevity of the presentation, we will focus on the case $S_{\delta,\intn}=\mathbbl{0}^\mathsf{T}_{n_{\mc N} - n_{\mc N^\ext}}$ in the remainder of this manuscript.

\begin{lemma}{\bf (Rank of $\boldsymbol{J}_\intn$)}\label{lemma:rankJint}
 If the network is connected and {interior-exterior node connected}, then $J_\intn$ has full rank.
\end{lemma}
\begin{IEEEproof}
By reordering its columns, the incidence matrix can be partitioned into
\begin{align}
 B = \begin{bmatrix} B^\ext_{\ext} & B^\ext_\text{c} & \mathbbl{0}_{n_{\mc N^\ext}  \times n_{\mc E^\intn}} \\
               \mathbbl{0}_{n_{\mc N^\intn} \times n_{\mc E^\ext}}  & B^\intn_\text{c} & B^\intn_\intn
     \end{bmatrix},
\end{align}
where $B^\ext_{\ext}$ models branches between external nodes, $B_c=[(B^\ext_\text{c})^\mathsf{T} \; (B^\intn_\text{c})^\mathsf{T}]^\mathsf{T}$ models branches between external and internal nodes, and $B^\intn_\intn$ models branches between internal nodes. Partitioning $W$ accordingly (i.e., $W=\diag\{W_\ext,W_c,W_\intn\}$) and noting that $W_c \succ 0$ and $W_\intn \succ 0$), it follows that $J_\intn = [B^\intn_\text{c} \; B^\intn_\intn] \diag\{W_c,W_\intn\} [B^\intn_\text{c} \; B^\intn_\intn]^\mathsf{T}$. Thus, it remains to show that $[B^\intn_\text{c} \; B^\intn_\intn]$ has full rank. To this end, the block columns $B^\intn_{\text{c},l}$ and $B^\intn_{\intn,l}$ of $B^\intn_{\text{c}}$ and $B^\intn_{\intn}$ can be further partitioned into
\begin{align*}
B^\intn_{\text{c},l} = \begin{bmatrix} B^\intn_{\text{c},l,|\mc N^\intn|+1} \\ \vdots \\ B^\intn_{\text{c},l,|\mc N|}\end{bmatrix},\quad B^\intn_{\intn,l} = \begin{bmatrix} B^\intn_{\intn,l,|\mc N^\intn|+1} \\ \vdots \\ B^\intn_{\intn,l,|\mc N|}\end{bmatrix}.
\end{align*}
For all $(l_i,l_k) \in \mc E_\sync \cap (\mc N^\intn \times \mc N^\ext)$ the only non-zero block of $B^\intn_{\text{c},l}$ is $B^\intn_{\text{c},l,l_i}=\mathsf{J}_{l_i l_i} =I_{2 n_{l_i}}$ and $B^\intn_{\text{c},l}$ has full column rank. Similarly, for all $(l_i,l_k) \in \mc E_\sync \cap (\mc N^\intn \times \mc N^\intn)$, the only non-zero blocks of $B^\intn_{\intn,l}$ are $B^\intn_{\intn,l,l_i}=\mathsf{J}_{l_i l_i}=I_{2 n_{l_i}}$, $B^\intn_{\intn,l,l_k} = \mathsf{J}_{\mathsf{l_i l_k}}$, and $B^\intn_{\intn,l}$ has full column rank. Moreover, $B^\intn_{\text{c},l}$ and $B^\intn_{\text{c},r}$ have linearly independent columns for all $l_i \neq r_i$, and $B^\intn_{\intn,l}$ and $B^\intn_{\intn,r}$ have linearly independent columns $l_i \neq r_i$.

Finally, by Assumption~\ref{assum:wp}.\ref{assum:wp:1} and Definition~\ref{def:1}, for every interior node $k \in \mc N^\intn$, there exists at least one block column $B^{\intn\star}_k=B^\intn_{\text{c},l}$ or $B^{\intn\star}_k=B^\intn_{\intn,l}$ with $l_k=k$ and $(l_i,l_k) \in \mc E_\sync \cap (\mc N^\intn \times \mc N)$. It directly follows that $\rank([B^\intn_\text{c} \; B^\intn_\intn]) \geq \rank([B^{\intn\star}_{|\mc N^\ext|+1} \; \ldots \; B^{\intn\star}_{|\mc N|}]) = n_{\mc N^\intn}$, i.e., $J_\intn$ has full rank.
\end{IEEEproof}
We emphasize that the proof crucially hinges on the fact that the network is connected and {interior-exterior node connected} network, i.e., every interior node is connected to a terminal with $\mathsf{J}_\mathsf{ii}=I_6$ or $\mathsf{J}_\mathsf{ii}=I_2$. In contrast, one can easily construct networks that are not {interior-exterior node connected} according to Definition~\ref{def:1} for which $J_\intn$ is not full rank.

Using Lemma~\ref{lemma:rankJint} and \eqref{eq:networkpowerflow}, the interior voltages of an {interior-exterior node connected} network satisfy  $V_{\delta}^\ext=J_\intn^{-1} S_{\delta,\intn}-J_\intn^{-1} J_\text{c}^\mathsf{T}  V_\delta^\ext$. This results in the reduced network model
\begin{align}\label{eq:rednet}
 S_{\delta,\ext} = (J_\ext-J_\text{c} J_\intn^{-1} J_\text{c}^\mathsf{T}) V_\delta^\ext + J_\text{c} J_\intn^{-1} S_{\delta,\intn}.
\end{align}
Thus, Lemma~\ref{lemma:rankJint} provides conditions under which Kron reduction (see e.g., \cite{dorfler2012kron}) can be applied to quasi-steady-state models of unbalanced power networks with common three-phase and single-phase branches (i.e., lines and transformers). In the remainder, we will focus on the case $S_{\delta,\intn}=\mathbbl{0}^\mathsf{T}_{n_{\mc N} - n_{\mc N^\ext}}$.

\section{Grid-forming power converter control}\label{Droop}
In our setting, the three-phase exterior voltages $V_{\delta,i}$, $i\in \mc N^{\ext}_{3\phi}$ are either imposed by a three-phase VSC or three ($\Wyegnd$-connected) single-phase VSCs. {We first discuss how to model standard single-phase and three-phase grid-forming droop control in our modeling framework and then introduce the novel phase-balancing feedback.}

{\subsection{Single-phase and three-phase converter control}
For single-phase VSCs (i.e., $i\in \mc N^{\ext}_{1\phi}$), single-phase droop control \cite{TJU+1997}, i.e., 
\begin{subequations}
\begin{align}\label{eq:droop_1phi_orig}
  \ddt \theta_i &= \omega_0 + m_p (P^\star_i - P_i),\\
 \tau \ddt V_i &= -V_i + V^\star_i + m_q (Q^\star_i - Q_i)
\end{align}
\end{subequations}
with droop coefficient $m_p \in \mathbb{R}_{>0}$ and $m_q \in \mathbb{R}_{>0}$ as well as low pass filter time constant $\tau \in {\mathbb{R}_{>0}}$ can be modeled in our framework by}
\begin{align}\label{eq:droop_1phi}
      \begin{bmatrix} \dot\theta_{\delta,i} \\ \dot v_{\delta,i}
   \end{bmatrix}
        \!=&\!
      -m_d \begin{bmatrix}
        P_{\delta,i} \\ Q_{\delta,i}
        \end{bmatrix}
        -\left(\frac{1}{\tau}\mathsf{F}_\mathsf{i} \mathsf{F}_\mathsf{i}^\mathsf{T}\right)  \begin{bmatrix} \theta_{\delta,i} \\  v_{\delta,i}
   \end{bmatrix},
\end{align}
and $\mathsf{F}_\mathsf{i}\coloneqq 	[\mathbbl{0}_{n_i \times n_i} \;  I_{n_i} ]^\mathsf{T}$. {In other words, $P_{\delta,i}$, $Q_{\delta,i}$, and $v_{\delta,i}$ denote deviations from the converter setpoints for active power, reactive power, and voltage magnitude. Notably, for brevity of the presentation, \eqref{eq:droop_1phi} assumes that $m_d = m_p = m_q / \tau$ is the normalized droop coefficient. 

Similarly, letting $\gamma_i \in \mathbb{R}$ and $\vartheta_i \in \mathbb{R}$ denote the  positive sequence voltage phase angle and magnitude deviation, standard three-phase droop control \cite{MCC-DMD-RA:93,RLB+2012,Guerrero2015,simpson2017voltage} for three-phase VSCs (i.e., $i\in \mc N^{\ext}_{3\phi}$) is modeled by $\mathsf{E} \coloneqq I_2 \otimes \mathbbl{1}_3$ and}
 \begin{align}\label{eq:posdroop}
 \begin{bmatrix} \theta_{\delta,i} \\ v_{\delta,i}
   \end{bmatrix}\! = \mathsf{E}\! \begin{bmatrix} \gamma_i \\ \vartheta_i \end{bmatrix}\!\! ,  \quad\!\! \!\begin{bmatrix} \dot \gamma_i \\ \dot \vartheta_i  \end{bmatrix}
\!=\! -m_d \mathsf{E}^\mathsf{T}\!\! \begin{bmatrix}
        P_{\delta,i} \\ Q_{\delta,i}
        \end{bmatrix}\!\! -\!\begin{bmatrix} 0 &  0 \\ 0 & \frac{1}{\tau} \end{bmatrix}\!\! \begin{bmatrix} \gamma_i \\ \vartheta_i  \end{bmatrix}\!\!.
\end{align}

For buses with three single-phase VSCs\footnote{We note that a neutral point clamped three-phase VSC can alternatively be modeled as three $\Wyegnd$-connected single-phase VSCs} we propose the novel generalized three-phase droop control
\begin{align} \label{eq:droop_3phi}
   \begin{bmatrix} \dot\theta_{\delta,i} \\ \dot v_{\delta,i}
   \!\!\end{bmatrix}
        \!=&\!
      -m_d
      \begin{bmatrix}
        P_{\delta,i} \\ Q_{\delta,i}
        \end{bmatrix}
        \!-\!\left(\frac{1}{\tau}\mathsf{F}_\mathsf{i} \mathsf{F}_\mathsf{i}^\mathsf{T}+k_{\text{\upshape{bal}},i} \mathsf{S}_\mathsf{i} \mathsf{S}_\mathsf{i}^\mathsf{T}\right)\!  \begin{bmatrix} \theta_{\delta,i} \\  v_{\delta,i},
\end{bmatrix},\!\!
\end{align}
i.e., single-phase droop control with $\mathsf{F}_\mathsf{i}$ and an additional {phase-balancing} feedback with gain $k_{\text{\upshape{bal}},i} \in \mathbb{R}_{\geq 0}$ and $\mathsf{S}_\mathsf{i} \coloneqq I_2 \otimes  \sqrt{12} \mathsf{P}_\mathsf{4}$. Notably, $\mathsf{S}_\mathsf{i} \mathsf{S}_\mathsf{i}^\mathsf{T}$ is a Laplacian matrix {and induces phase-balancing.}

{
\subsection{Control gains, setpoints, and implementation}
The droop coefficients $m_p \in \mathbb{R}_{>0}$ and $m_q \in \mathbb{R}_{>0}$ are typically in the range of $1-5\%$ and prescribed by system operators, grid-codes, or through markets. Moreover, $\tau$ models a low-pass filter time constant in the range from $20~\mathrm{ms}$ to $200~\mathrm{ms}$. In other words, in a large-scale power system $m_p$ and $m_q$ are typically not degrees of freedom for control design and typically assumed to be identical for each device in per unit (i.e., proportional to the rating of the converter).

Within the standard power system control hierarchy the controllers \eqref{eq:droop_1phi}, \eqref{eq:posdroop}, and \eqref{eq:droop_3phi} can be classified as primary control and can readily use standard dispatch signals from secondary and tertiary control layers. On the timescales of interest (milliseconds to seconds) the setpoints $(P^\star_i, Q^\star_i, V^\star_i$) from higher-level controls are commonly assumed to be constant. Moreover, in our small-signal model the active and reactive power setpoints are zero because the network model is linearized around the zero power flow solution.

Finally, we emphasize that, in contrast to works that leverage power flow optimization to mitigate voltage unbalances \cite{Line,Rahul,Arnold, su2014optimal, karagiannopoulos2018}, the phase-balancing feedback in \eqref{eq:droop_3phi} does not require a model of the network and only requires local communication. As we will show in the next section, certifying small-signal stability of the converter-dominated power system only requires knowledge of the network topology but not the exact line or transformer parameters.}

{\section{Stability conditions}\label{sec:stability}
In this section, we provide analytical stability conditions power systems containing a mix of three-phase and single-phase grid-forming converters. We first establish several preliminary properties of the overall system (i.e., grid-forming converters connected through the network) and establish analytical results for low-complexity network structures (e.g., radial network with only three-phase branches and one exterior bus). Subsequently, we extend these results to networks of increased complexity to arrive at stability conditions for a general network containing three-phase and single-phase networks and standard transformer interconnections. }

{\subsection{Overall system dynamics}}
{We first develop a model of the overall system obtained by interconnecting the grid-forming  controls \eqref{eq:droop_1phi}, \eqref{eq:posdroop}, and \eqref{eq:droop_3phi} through the reduced order network model \eqref{eq:rednet}. To this end,} we define the matrices $F_i \coloneqq [F_{i,1} \; \ldots \; F_{i,n_{|\mc N|}}] \in \mathbb{R}^{n_{\mc N} \times 2 n_i}$ and $S_i\coloneqq [S_{i,1} \; \ldots \; S_{i,n_{|\mc N|}}] \in \mathbb{R}^{n_{\mc N} \times 2 n_i}$ that collect the matrices $\mathsf{F}_\mathsf{i}$, i.e., with $F_{i,j} = \mathsf{F}_\mathsf{i}$ for all buses $i$ using \eqref{eq:droop_3phi} or \eqref{eq:droop_1phi} and $i=j${, with $S_{i,j} = \mathsf{S}_\mathsf{i}$ for all buses $i$ using \eqref{eq:droop_3phi}}, and $F_{i,j}$ and $S_{i,j}$ are zero otherwise. Similarly, we define $E=\diag\{E_i\}_{i=1}^{n_{|\mc N|}}$ with $E_i=\mathsf{E}_i$ for nodes $i$ using \eqref{eq:posdroop} and $E_i=I_{n_i}$ otherwise. This results in the closed-loop matrix 
\begin{align*}
 J_{\text{\upshape{cl}}}\coloneqq E^\mathsf{T} \left(m_d J + \sum\nolimits_{i\in \mc N^\ext} \frac{1}{\tau}F_i  F_i^\mathsf{T} +  
k_{\text{\upshape{bal}},i} S_i S^\mathsf{T}_i\right) E,
\end{align*}
of the network with controlled three-phase and single-phase converters. Using the same partitioning as in \eqref{eq:networkpowerflow} and eliminating the interior nodes results in the closed-loop dynamics
\begin{align}\label{eq:reddyn}
 \ddt x  = -(J_{\text{\upshape{cl}},\ext}-J_\text{\upshape{cl,c}} J_{\text{\upshape{cl}},\intn}^{-1} J_\text{\upshape{cl,c}}^\mathsf{T}) x = J_\text{\upshape{cl},\upshape{red}}  x
\end{align}
with state vector $x \!\coloneqq\!(x_1,\ldots,x_{|N^\ext|})$, i.e., $x_i\!=\!(\gamma_i, \vartheta_i)$ for VSCs using \eqref{eq:posdroop} and $x_i\!=\!(\theta_{\delta,i},v_{\delta,i})$ when using \eqref{eq:droop_3phi} or \eqref{eq:droop_1phi}.
{

\subsection{Preliminary results}
Next,} we characterize the nullspace of the closed-loop matrix $J_\text{\upshape{T,cl}}$ of an {interior-exterior node connected} radial three-phase network that only contains one exterior node, three-phase lines, $\Wyegnd\Delta$ transformers, and $\Wyegnd\Wyegnd$ transformers. To this end, we define 
\begin{align*}
 \mu_3\coloneqq(I_3,\mathbbl{0}_{3\times3}), \quad h \coloneqq \begin{bmatrix}
	\mathsf{P}_{\mathsf{3}}& \mathsf{P}_{\mathsf{4}}\\
	-\mathsf{P}_{\mathsf{4}}& \mathsf{P}_{\mathsf{3}}\end{bmatrix}.
\end{align*}
\begin{lemma}{\bf{(Nullspace of $\boldsymbol{J}_\text{\upshape{T,cl}}$)}}\label{lemma:2} 
Consider a radial three-phase network $\mc G_{\text{\upshape{T}}} \!=\! (\mc E_{\text{\upshape{T}}},\mc N_{\text{\upshape{T}}})$ {with $\mc N_{\text{\upshape{T}}}^\ext\!\coloneqq\!\{1\}$ and $\mc E_{\text{\upshape{T}}}\!=\! \mc E_{3\pi} \cup \mc E_{\Wyegnd\Wyegnd} \cup \mc E_{\Wyegnd\Delta}$} that satisfies Assumption~\ref{assum:wp}. Then, the following holds:
\begin{enumerate}
 \item Using \eqref{eq:posdroop}, $J_\text{\upshape{T,cl}} \xi = \mathbbl{0}_{6 |\mc N_\text{\upshape{T}}|-4}$ if and only if $\xi \in \vspan\big((\nu_1,\mathbbl{1}_{|\mc N_\text{\upshape{T}}|-1} \otimes \nu_3)\big)$,
 
 \item Using \eqref{eq:droop_3phi} with $k_{\text{\upshape{bal}},1} \in \mathbb{R}_{>0}$, $J_\text{\upshape{T,cl}} \xi = \mathbbl{0}_{6 |\mc N_\text{\upshape{T}}|}$ if and only if $\xi \in \vspan(\mathbbl{1}_{|\mc N_\text{\upshape{T}}|} \otimes \nu_3)$,
 
 \item Using \eqref{eq:droop_3phi} with $k_{\text{\upshape{bal}},1}\!=\!0$, let $r_p$ denotes the number of $\Wyegnd\Delta$ edges between $p \in \mc N^\intn$ and the exterior node, then $J_\text{\upshape{T,cl}} \xi = \mathbbl{0}_{6 |\mc N_\text{\upshape{T}}|}$ if and only if $\xi_p \in \vspan(h^{r_p} \mu_3)$.
 
\end{enumerate}
\end{lemma}
\begin{IEEEproof}
The closed-loop network matrix of $\mc G_\text{T}$ is given by
\begin{align} \label{eq:12}
    J_{\text{T,cl}}\coloneqq E^\mathsf{T}\bigg(m_d J_\text{T} + \frac{1}{\tau}  F_1  F_1^\mathsf{T} +  
k_{\text{\upshape{bal}},1} S_1 S^\mathsf{T}_1\bigg) E.
\end{align}
Because $B_l B_l^\mathsf{T} \succcurlyeq 0$, $F_1 F_1^\mathsf{T} \succcurlyeq 0$, and $S_1 S_1^\mathsf{T} \succcurlyeq 0$ it follows that ${J}_\text{T,cl} \xi = \mathbbl{0}_{6 |\mc N_\text{T}|}$ if and only if $B_l^\mathsf{T} \xi = \mathbbl{0}_{6 |\mc N_\text{T}|}$,  for all $l \in \mathbb{N}_{[1,|\mc E_\text{T}|]}$, and $F_1^\mathsf{T} \xi = \mathbbl{0}_{6 |\mc N_\text{T}|}$ as well as $S_1^\mathsf{T} \xi = \mathbbl{0}_{6 |\mc N_\text{T}|}$. Moreover, for any edge $l \in \mathbb{N}_{[1,|\mc E_\text{T}|]}$ with $(l_i,l_k) \in \mc E_\text{T}= \mc E_{3\pi} \cup \mc E_{\Wyegnd\Wyegnd} \cup \mc E_{\Wyegnd\Delta}$, $B_l \xi = \mathbbl{0}_{6 |\mc N_\text{T}|}$ implies $\xi_{l_k}= {-\mathsf{J^{\mathsf{-T}}_{l_i l_k}} \mathsf{J^{\mathsf{T}}_{l_i l_i}}} \xi_{l_i}$. 

Next, assume that the exterior node uses \eqref{eq:droop_3phi} and  $k_{\text{\upshape{bal}},1} \in \mathbb{R}_{>0}$. Then $F_1 \xi =\mathbbl{0}_{6 |\mc N_\text{T}|}$ and $S_1 \xi = \mathbbl{0}_{6 |\mc N_\text{T}|}$ if and only if $\xi_{1} \in \vspan(\nu_3)$. Moreover, for all $(l_i,l_k) \in \mc E_\text{T}= \mc E_{3\pi} \cup \mc E_{\Wyegnd\Wyegnd} \cup \mc E_{\Wyegnd\Delta}$ it holds that {$-\mathsf{J^{\mathsf{-T}}_{l_i l_k}} \mathsf{J^{\mathsf{T}}_{l_i l_i}} \nu_3 = \nu_3$}. It follows from induction over all edges in $\mc E_\text{T}$ that $\xi_{i}\in \vspan(\nu_3)$ needs to hold for all $i \in \mc N^\intn$. 

In contrast, using \eqref{eq:posdroop}, $J_\text{\upshape{T,cl}} \xi = \mathbbl{0}_{6 |\mc N_\text{\upshape{T}}|-4}$ if and only if $\xi_1 \in \vspan(\nu_1)$. Moreover, for all $(l_1,l_k) \in \mc E_\text{T}= \mc E_{3\pi} \cup \mc E_{\Wyegnd\Wyegnd} \cup \mc E_{\Wyegnd\Delta}$ it holds that $\xi_k = -\mathsf{J^{\mathsf{-T}}_{l_1 l_k}} \mathsf{J^{\mathsf{T}}_{l_1 l_1}} E \nu_1 = \nu_3$. It again follows from induction over all edges in $\mc E_\text{T}$ that $\xi_{i}\in \vspan(\nu_3)$ needs to hold for all $i \in \mc N^\intn \setminus \{1\}$. 

On the other hand, if $k_{\text{\upshape{bal}},1} =0$, then $F_1 \xi =\mathbbl{0}_{6 |\mc N_\text{T}|}$ and $S_1 \xi =\mathbbl{0}_{6 |\mc N_\text{T}|}$ imply $\xi_{1} \in \vspan(\mu_3)$. For all $(l_i,l_k) \in \mc E_\text{T}= \mc E_{3\pi} \cup \mc E_{\Wyegnd\Wyegnd}$ it holds that $-\mathsf{J^{\mathsf{-T}}_{l_i l_k}} \mathsf{J^{\mathsf{T}}_{l_i l_i}}=I_3$. Moreover, we note that $h$ is periodic (i.e., $h^7=h$) and has full rank. Next, for any interior node $p \in \mc N^\intn_\text{T}$, we use $r_{p}$ to denote the number of $\Wyegnd\Delta$ edges in the path between the node $p$ and the exterior node $1$. Then, it has to hold that $\xi_{p} \in \vspan(h^{r_p} \mu_3)$. Thus, if $k_{\text{\upshape{bal}},1} =0$ the nullspace of $J_\text{T,cl}$ has dimension three.
\end{IEEEproof}

The proof of Lemma~\ref{lemma:2} highlights that, up to modifying the dimension of $J_\text{cl}$ and its null vector, three-phase droop control \eqref{eq:posdroop} and generalized three-phase droop control \eqref{eq:droop_3phi} with $k_{\text{\upshape{bal}},i} \in \mathbb{R}_{>0}$ are interchangeable when analyzing the nullspace of $J_\text{cl}$. Thus, for brevity of the presentation we will focus on the controls \eqref{eq:droop_3phi} and \eqref{eq:droop_1phi} and only revisit the control \eqref{eq:posdroop} at the end of this section. Next, we generalize the results of Lemma~\ref{lemma:2} to two {interior-exterior node connected} radial three-phase networks connected via a three-phase branch.
\begin{lemma}{\bf{(Nullspace of two {interior-exterior node connected} three-phase networks)}}\label{lemma:3}
Consider two radial three-phase networks $\mc G_{\text{\upshape{T}}_1} = (\mc E_{\text{\upshape{T}}_k},\mc N_{\text{\upshape{T}}_k})$, $k\in\{1,2\}$ with $\mc E_{\text{\upshape{T}}_k}= \mc E_{3\pi} \cup \mc E_{\Wyegnd\Wyegnd} \cup \mc E_{\Wyegnd\Delta}$ and $\mc N^\ext_{\text{\upshape{T}}_k} \coloneqq\{1\}$ that satisfy Assumption~\ref{assum:wp} and are connected via $(l_i,l_j)\in\mc E_{3\phi}$. Moreover, one of the following holds:
\begin{enumerate}
 \item $k_{\text{\upshape{bal}},1} \in \mathbb{R}_{>0}$ or $k_{\text{\upshape{bal}},2} \in \mathbb{R}_{>0}$,
 \item there exists at least one path between the two exterior nodes that contains at least one $\Wyegnd\Delta$ branch and all $\Wyegnd\Delta$ branches are traversed in the same orientations.
\end{enumerate}
Then $J_{\text{\upshape{T}}_k,\text{\upshape{cl}}} \xi = \mathbbl{0}_{6 |\mc N_{\text{\upshape{T}}_k}|}$ if and only if $\xi \in \vspan(\mathbbl{1}_{|\mc N_{\text{\upshape{T}}_k}|} \otimes \nu_3)$.
 \end{lemma}

\begin{IEEEproof}
consider two radially three-phase networks $\mc G_{\text{\upshape{T}}_1}$ and $\mc G_{\text{\upshape{T}}_2}$ each with one exterior bus with VSC using generalized three-phase droop control and $\mc E_{\text{\upshape{T}}_r} \subseteq \mc E_{3\pi} \cup \mc E_{\Wyegnd\Wyegnd} \cup \mc E_{\Wyegnd\Delta}$ for $r \in \{1,2\}$. Assume these two networks are connected with an edge $(l_i,l_j)\in\mc E_{3\phi}$ 
(e.g,. see Fig.~\ref{TwoRadialNetwork}).
\begin{figure}[t]
\centering
\includegraphics[width=0.35\textwidth]{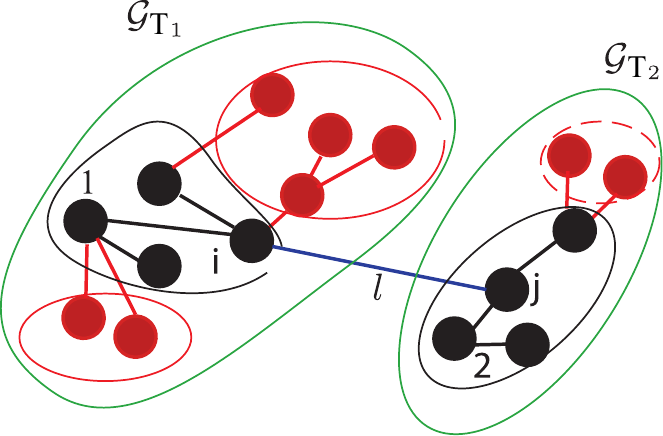}
\caption{Two radial networks that only contains three-phase lines and buses (black), single-phase lines and buses (red), and $\Wyegnd\Delta$ transformers. Each network has one exterior bus (bus 1 and 2) with generalized three-phase droop control. These networks are connected via a three-phase edge (blue).\label{TwoRadialNetwork}}
\end{figure}
Without loss of generality, we assume $k_{\text{\upshape{bal}},1} \in \mathbb{R}_{>0}$. Then Lemma~\ref{lemma:2} shows $J_{\text{\upshape{T}}_1,\text{\upshape{cl}}} \xi = \mathbbl{0}_{6 |\mc N_{\text{\upshape{T}}_1}|}$ if and only if $\xi \in \vspan(\mathbbl{1}_{|\mc N_{\text{\upshape{T}}_1}|} \otimes \nu_3)$. Moreover, for any $j\in \mc N_{\text{\upshape{T}}_2}$, it holds that $\xi_j \in\vspan\left(h^{r_j}\mu_3\right)$, where $r_{j}$ is the number of $\Wyegnd\Delta$ edges between the node $j$ and the exterior bus $\mc N_{\text{\upshape{T}}_2}$. Thus, every branch must satisfy
\begin{align} \label{eq:combined_network}
\begin{bmatrix}
	\mathsf{J}_{\mathsf{ii}}^\mathsf{T} & \mathsf{J}_{\mathsf{ij}}^\mathsf{T}
\end{bmatrix}
\begin{bmatrix}
	\nu_3\\
	h^{r_{j}}\mu_3\alpha 
\end{bmatrix}= 
	\mathbbl{0}_{6}
\end{align}
for some $\alpha \in \mathbbl{R}^3$. We note that $h=h^7$, has full rank, and \eqref{eq:combined_network} has the unique solution $\gamma=\mathbbl{1}_3$ for all $r_j \in \mathbb{N}_{[1,7]}$. Then, the first part of the lemma immediately follows from $h^{r_j}\mu_3\gamma=\nu_3$.
Next, assume that $k_{\text{\upshape{bal}},1}=k_{\text{\upshape{bal}},2}=0$ but that there exists at least one path between the two exterior nodes that contains at least one $\Wyegnd\Delta$ branch and all $\Wyegnd\Delta$ branches are traversed in the same orientations. Moreover, let $r_1$ denote the number of $\Wyegnd\Delta$ edges between the two exterior nodes.
Then, $\xi_1$ needs to be in the nullspace of $\mathsf{F}_1$, i.e., the last three rows of the vector $\xi_1$ are all zero, and \eqref{eq:combined_network} becomes $[\mathbbl{0}_{3\times3} \; I_3 ] h^{r_1}\mu_3\gamma= \mathbbl{0}_{3}$ and has the unique solution $\gamma=\mathbbl{1}_3$ for $r_1\in \mathbb{N}_{[1,7]}$. The second part of the lemma directly follows from $h^{r_1}\mu_3\gamma=\nu_3$.
\end{IEEEproof}
Next, we extend the results of Lemma~\ref{lemma:3} to a general connected and {interior-exterior node connected} network.
\begin{theorem}
{\bf{(Nullspace of $\boldsymbol{J_\text{cl}}$)}}\label{Theorem:1}
Consider a connected network $\mc G = (\mc E,\mc N)$ that satisfies  Assumption~\ref{assum:wp} and VSCs using \eqref{eq:droop_3phi} and \eqref{eq:droop_1phi}. Assume that one of the following holds:
 \begin{enumerate}
 \item there exists $i\in \mc N^{\ext}_{3\phi}$ such that $k_{\text{\upshape{bal}},i} \in \mathbb{R}_{>0}$,
 \item there exists at least one path between two exterior nodes that contains at least one $\Wyegnd\Delta$ branch and all $\Wyegnd\Delta$ branches are traversed in the same orientation.
\end{enumerate}
 Then $J_\text{\upshape{cl}} \xi = \mathbbl{0}_{6 |\mc N_{3\phi}| + 2 |\mc N_{1\phi}|}$ if and only if $\xi_i \in \vspan(\nu_{n_i})$.
 \end{theorem}
\begin{IEEEproof}
Any network that satisfies Assumption~\ref{assum:wp} can be decomposed into (i) $p$ {interior-exterior node connected}, connected, and radial three-phase networks $\mc G_{\text{\upshape{T}}_r}=(\mc E_{\text{\upshape{T}}_r},\mc N_{\text{\upshape{T}}_r})$ such that $\mc E_{\text{\upshape{T}}_r} \subseteq \mc E_{3\pi} \cup \mc E_{\Wyegnd\Wyegnd} \cup \mc E_{\Wyegnd\Delta}$ and $\cup_{r=1}^p \mc N_{\text{\upshape{T}}_r} = \mc N_{3\phi}$, (ii) three-phase branches $\mc E^\prime_{3\phi} \coloneqq \mc E_{3\phi} \setminus \cup_{r=1}^p  \mc E_{\text{\upshape{T}}_r}$ connecting the networks $G_{\text{\upshape{T}}_r}$, and (3) $\kappa$ single phase networks that are each connected to exactly one three-phase bus. Thus, the closed-loop network matrix of an arbitrary network can be expressed as
\begin{align*}
    J_\text{\upshape{cl}}\coloneqq \sum_{r=1}^p {J}_{\text{T}_{r},\text{cl}} + \sum_{(l_i,l_j)\in \mc{E}^\prime_{3\phi}} \mc J_l + \sum_{r=1}^\kappa {J}_{{1\phi}_r,\text{cl}}    + \Xi,
\end{align*}
where the remaining three-phase edges and matrices modeling the controls of three-phase converters that are not part of the $p$ {interior-exterior node connected} radial three-phase networks or their interconnection are collected in the matrix $\Xi$. Note that $\Xi \succeq 0$ does not increase the dimension of the nullspace of $J_\text{\upshape{cl}}$ and can be neglected in the following arguments.

Next, under the hypothesis of the theorem, it follows from Lemma~\ref{lemma:2} that there exists $r \in \{1,\ldots,p\}$ such that $\ker(J_{\text{T}_r,\text{cl}})=\vspan(\mathbbl{1}_{|\mc N_{\text{\upshape{T}}_r}|} \otimes \nu_3)$. Repeated application of  Lemma~\ref{lemma:3} extends this result to all $r \in \{1,\ldots,p\}$. In addition, any vector in the nullspace of $J_\text{\upshape{cl}}$ must also be in the nullspace of $F_i  F_i^\mathsf{T}\succcurlyeq 0$, and $\mathcal{J}_{(l_i,l_j)\in \mc{E}_{1\phi}}\succcurlyeq 0$. In other words, $J_\text{\upshape{cl}} \xi = \mathbbl{0}_{6 |\mc N_{3\phi}| + 2 |\mc N_{1\phi}|}$ if and only if $\xi_i \in \vspan(\nu_{n_i})$ for all $i \in \mc N_{1\phi}$ and the theorem directly follows.
\end{IEEEproof}

{\subsection{Main result}}
We can now use Theorem~\ref{Theorem:1} to show that, under mild conditions, the dynamics \eqref{eq:reddyn} are asymptotically stable with respect to $x_i = \nu_{n_i}$ for all $i\in\mc N^{\ext}$, i.e., identical voltage phase angle deviations $\theta_{\delta,i}$ and zero voltage magnitude deviation $v_\delta$. {In other words, the following Theorem extends dynamic stability results to power systems with standard transformer interconnections and a mix of grid-forming converters using (i) single-phase droop control \eqref{eq:droop_1phi}, (ii) standard three-phase droop control \eqref{eq:posdroop}, and (iii) generalized three-phase droop control \eqref{eq:droop_3phi}.}

\begin{theorem}\label{thm:main}{\bf (Asymptotic stability) }\label{Theorem:2} Consider a {interior-exterior node connected network that satisfies Assumption~\ref{assum:wp}. Moreover, one of the following holds:
\begin{enumerate}
 \item there exists a three-phase VSC $i\in \mc N^{\ext}_{3\phi}$ using standard three-phase droop control \eqref{eq:posdroop},
 \item there exists a three-phase VSC $i\in \mc N^{\ext}_{3\phi}$ using generalized three-phase droop control \eqref{eq:droop_3phi} with $k_{\text{\upshape{bal}},i} \in \mathbb{R}_{>0}$,
 \item there exists at least one path between the two exterior nodes that contains at least one $\Wyegnd\Delta$ branch and all $\Wyegnd\Delta$ branches are traversed in the same orientations. 
\end{enumerate}}
Then, \eqref{eq:reddyn} is asymptotically stable with respect to the subspace $x_i = \nu_{n_i}$ for all $i\in\mc N^{\ext}$.
\end{theorem}
\begin{IEEEproof}
We first note that $J_\text{\upshape{cl,int}} = J_\text{\upshape{int}}$, i.e., by Lemma~\ref{lemma:rankJint} the inverse $\mc (J_\text{\upshape{cl,int}})^{-1}$ exists. {Next, we proof the Theorem under the hypothesis of item 2) and 3). By Theorem~\ref{Theorem:1}, $J_\text{\upshape{cl}} \xi = \mathbbl{0}_{6 |\mc N_{3\phi}| + 2 |\mc N_{1\phi}|}$ if and only if $\xi_i \in \vspan(\nu_{n_i})$. Then,} by the closure property of the Schur complement $J_\text{\upshape{cl},\upshape{red}}$ is negative semi-definite, the nullspace of $J_\text{\upshape{cl},\upshape{red}}$ and $J_\text{\upshape{cl}}$ has the same dimension, and the only vector contained in the nullspace of $J_\text{\upshape{cl},\upshape{red}}$ is given by $V_{\delta,i} \in \vspan((\mathbbl{1}_{n_i},\mathbbl{0}_{n_i}))$. 
{In other words, all but one eigenvalue of $J_\text{\upshape{cl},\upshape{red}}$ have negative real part, the remaining eigenvalue is zero, the eigenvector associated with the eigenvalue at zero is given by $V_{\delta,i} \in \vspan((\mathbbl{1}_{n_i},\mathbbl{0}_{n_i}))$, and the Theorem directly follows for item 2) and 3).}

{The proof of the Theorem under the hypothesis of item 1) is a direct consequence of the fact that the steps used to proof Lemma~\ref{lemma:3}, Theorem~\ref{Theorem:1}, and item 2) are applicable if any VSC using \eqref{eq:droop_3phi} is replaced with a VSC using \eqref{eq:posdroop}. The only difference is the number of states associated with the corresponding three-phase converter, i.e., the asymptotically stable solution is given by $x_i = \nu_1$ instead of $x_i=\nu_3$ if \eqref{eq:posdroop} is used at node $i\in\mc N^{\ext}$.}
\end{IEEEproof}

{
\subsection{Interpretation of the stability conditions} 
We first note that Theorem~\ref{Theorem:2} establishes stability of the converter-dominated power system for any positive normalized droop coefficient $m_d$, i.e., is in line with the paradigm that $m_p$ and $m_q$ are not a tuning parameter to ensure dynamic stability but specified by e.g., grid-codes or system operators based on steady-state considerations. Nonetheless, we note that the range of $m_p$, $m_q$, and $\tau$ has to be restricted to guarantee stability when dynamics of network branches (i.e., lines and transformers) are accounted for \cite{Guerrero2015,vorobev2017high,line_dynamics}. Typically, this effect is only relevant for large resistance to reactance ratios or very short lines \cite{vorobev2017high,line_dynamics}. A thorough investigation of this aspect is seen as an interesting topic for future research.

The conditions of Theorem~\ref{Theorem:2} can be understood as restricting the network topology to guarantee stability in three broad scenarios that we will discuss next. 

\subsubsection{Self-balancing of single-phase converters in a three-phase network}
In the most extreme case one may consider a three-phase system that only contains grid-forming single-phase converters. In such a system, the single-phase converters self-synchronize through the network power flows and phase-balance {\it{without any controller}}, if there is a path between the two exterior VSCs that contains at least one $\Wyegnd\Delta$ branch and all $\Wyegnd\Delta$ branches are traversed in the same orientations. This result generalized and formally explains the observation in \cite{Brian_Delta} that three (delta connected) single-phase converters connected to a three-phase may autonomously phase-balance. However, our results show that, in general autonomous phase-balancing of single-phase converters may not be a generic effect in three-phase networks and may strongly hinge on the network topology. Moreover, in this setup the trade-off between power and voltage unbalance between phases in the presence of unbalanced load is purely determined by the network parameters and loads and cannot be tuned.

\subsubsection{Synchronization of single-phase converters to three-phase converters}
On the other hand, Theorem~\ref{Theorem:2} shows that, under a wide range of network topologies, grid-forming single-phase converters implicitly phase-balance through synchronization with a grid-forming three-phase converter that imposes a balanced voltage at its terminals (i.e., uses standard droop control \eqref{eq:posdroop}). While sufficient from a theoretical point view, in practice one would not rely on a single device for system stability and synchronization and instead leverage multiple devices that satisfy the stability condition. Moreover, in this setup the trade-off between power and voltage unbalance between phases in the presence of unbalanced load is again purely determined by the network parameters.

\subsubsection{Single-phase using phase-balancing feedback}
In either of the previous cases the phase-balancing feedback in \eqref{eq:droop_3phi} can be applied to single-phase converters connected to the same bus to control the trade-off between voltage and power unbalance between phases in the presence of unbalanced load. In other words, voltage unbalance can be reduced at the expense of local communication. Moreover, compared to autonomous phase-balancing of single-phase converters, phase-balancing through feedback for at least one bus allows certifying stability for a wider class of network topologies. Investigating balancing feedback between single-phase converters that are not located at the same bus and clarifying the communication requirements are the focus of ongoing work.

Finally, we note that \eqref{eq:droop_3phi} can also be applied on three-phase converters to control the trade-off between voltage and power unbalance between phases in the presence of unbalanced load. In this case, the phase-balancing feedback is implemented within the same controller and does not require communication.}

\section{Case Studies}\label{Simulation}
Finally, we use an EMT simulation in MATLAB/Simulink to illustrate and validate the results. 

\begin{figure}[t!!!]
  \centering
  \includegraphics[width=0.9\columnwidth]{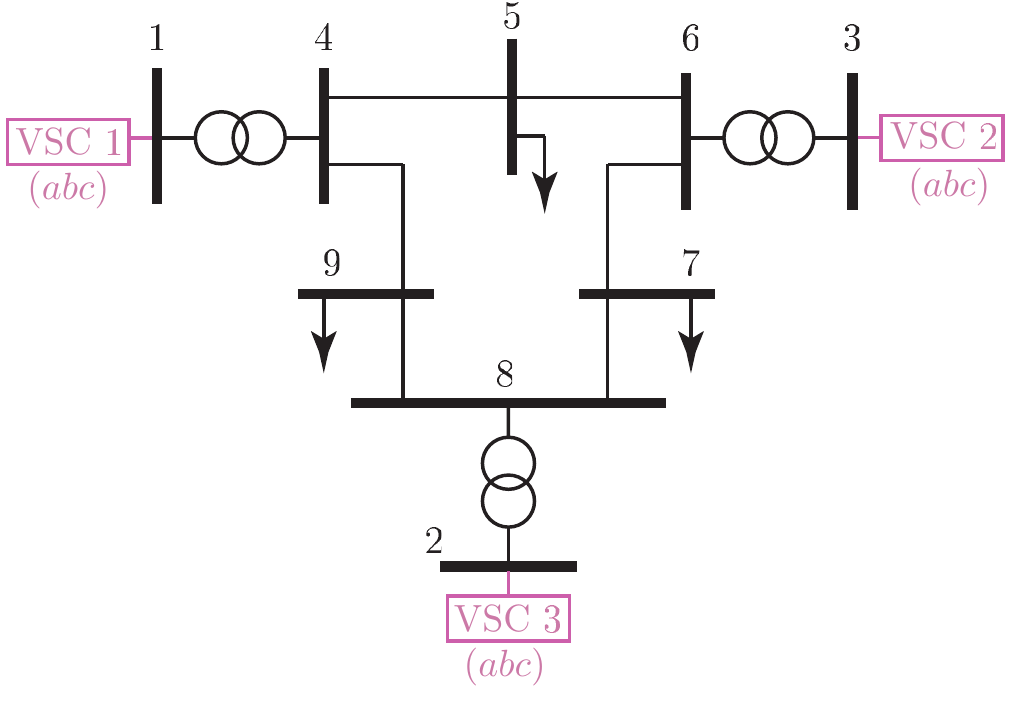}
  \caption{IEEE~9-bus transmission system. The three-phase VSCs use the standard positive sequence droop control, in which the voltage and angle corrections for all three phases in each VSCs are the same.\label{fig6}}
  \end{figure}

\begin{figure}[t!!!]
  \includegraphics[width=0.9\columnwidth]{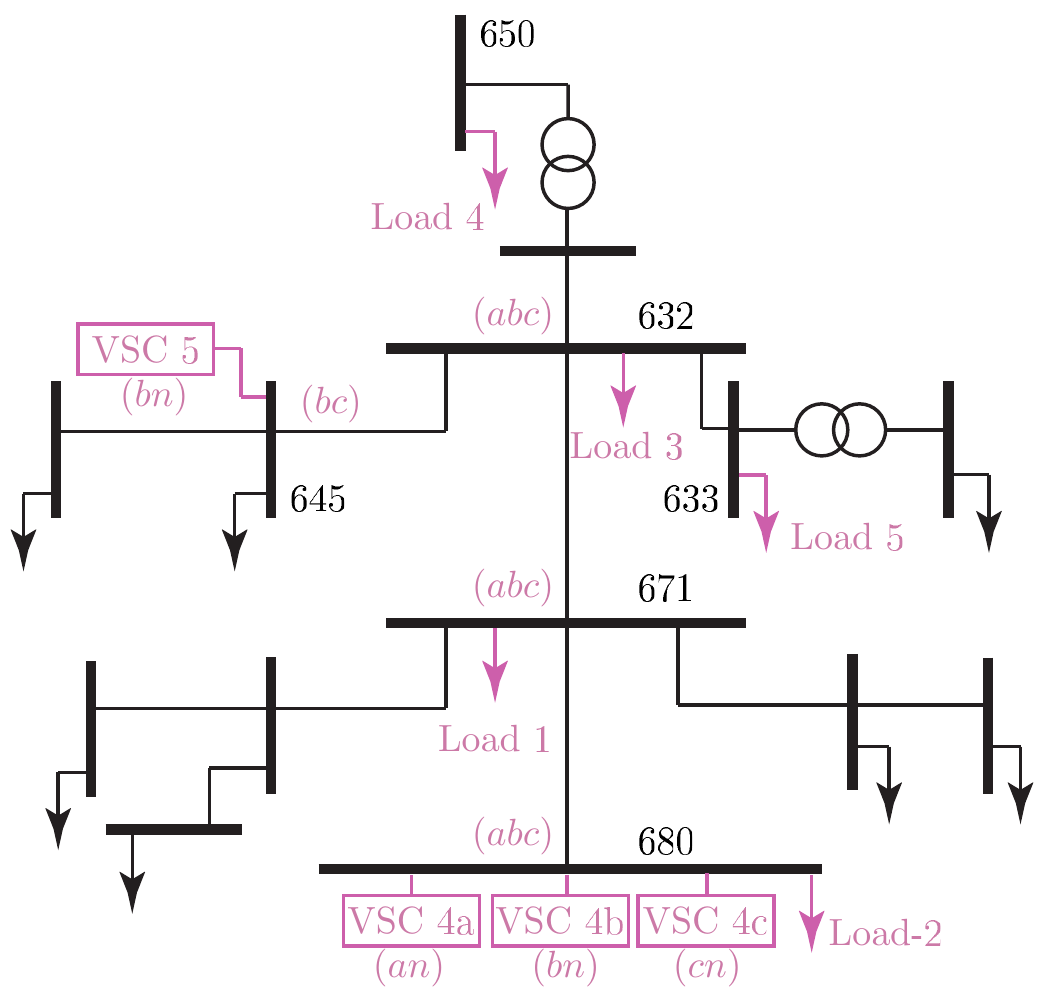}
  \caption{IEEE~13-bus distribution feeder with five delta connected constant impedance Loads (i.e., $\mathrm{Load~1}-\mathrm{Load~5}$). Single-phase VSCs at bus 680 and 645 are used for power generations.\label{fig7}}
  \end{figure}

\begin{figure}[t!!]
    \centering
    \includegraphics[width=0.5\columnwidth]{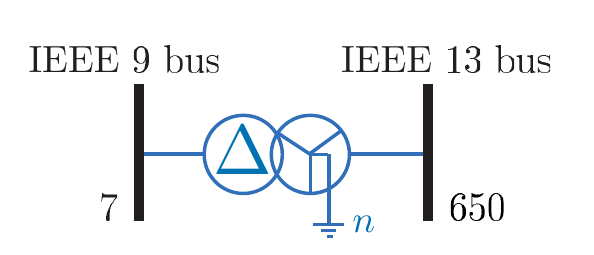}
    \caption{In the first case study, bus 680 of the IEEE~13-bus system is connected to the IEEE~9-bus system at bus 7. \label{fig71}}
    \end{figure}

{\subsection{Description of the system under study}
Our case study combines a three-phase (three-wire) transmission system (IEEE~9-bus with $100~\mathrm{MVA}$ base power and $230~\mathrm{kV}$ base voltage) and a four-wire  bus distribution system (IEEE~13-bus with $1.67~\mathrm{MVA}$ base power and $4.4~\mathrm{kV}$ base voltage). The parameters of the distribution system neutral lines are taken from data sheets for the type of line specified in the IEEE~13-bus system. We consider two cases. 

\subsubsection{Transmission and distribution connected VSCs} The first case represents a transition scenario to converter-dominated systems in which grid-forming VSCs are predominantly deployed at the transmission level. In this case, all synchronous generators in the IEEE~9-bus system are replaced by three-phase two-level voltage source converters (VSC) as shown in Fig.~\ref{fig6}). The VSC parameters are taken from \cite{tayyebi2020frequency} and the VSC ac voltage phase angles and magnitudes are determined using the standard three-phase droop control \eqref{eq:posdroop}. Moreover, one IEEE~13-bus system is connected to the IEEE~9-bus system through a step-up $\Delta\Wyegnd$ transformer (see Fig.~\ref{fig71}).
The IEEE~13-bus system contains four single-phase two-level VSCs with ac voltage phase angle and magnitude determined by the control \eqref{eq:droop_1phi}. Specifically,  single-phase VSCs are connected between each phase and neutral at bus 680, and a single-phase VSC is connected between phase b and neutral at bus 645. We also added five constant power loads to the buses 632, 633, 650, 671, and 680 (see Fig.~\ref{fig7}). All VSCs use an active power droop gain $m_p$ of $2\%$ and reactive power droop gain $m_q$ of $1\%$ (in system base) for active power and reactive power and a time constant of $\tau=1~\mathrm{ms}$ is used for all the simulations. Aggregated loads at buses 5, 7 and 9, are used to model, in abstraction, distribution feeders without converter-interfaced generation.

\subsubsection{Distribution connected single-phase VSCs} The second case represents an extreme scenario in which all generation is interfaced to the distribution system through single-phase VSCs. In this case, the synchronous generators in the original IEEE~9-bus transmission system at bus 1,2, and 3 are each replaced by an IEEE~13-bus system connected through a step-up $\Delta\Wyegnd$ transformers (see Fig.~\ref{fig72}). The IEEE~13-bus system again contains four single-phase two-level VSCs as in Case~1. In this case, the VSCs at bus 680 are controlled using the generalized three-phase droop control \eqref{eq:droop_3phi}, i.e., using phase-balancing feedback. Moreover, the aggregated loads at buses 5, 7 and 9, remain in the model to model, in abstraction, distribution feeders without converter-interfaced generation. Aggregated loads at buses 5, 7 and 9, are again used to model distribution feeders without converter-interfaced generation.}


Before presenting the simulation results, we introduce the following metrics for measuring unbalance.
\begin{definition} {\bf(Unbalance Factors)}\label{def:UF}
The voltage unbalance factor $V_{\text{\upshape{UF}},i}$ of a bus voltage phasor $V_i =(V_{a,i},V_{b,i},V_{c,i}) \in \mathbb{C}^3$ is commonly defined by $\mathsf{a}=e^{\frac{2}{3}j}$ and \cite{standard2002}
\begin{align*}
V_{\text{\upshape{UF}},i}=\frac{|V_{a,i}+ \mathsf{a}^2 V_{b,i}+ \mathsf{a} V_{c,i}|}{|V_{a,i} + \mathsf{a} V_{b,i} + \mathsf{a}^2 V_{c,i}|}.
\end{align*}
Moreover, for a three-phase node $i \in \mc N_{3\phi}$, we define the power unbalance {relative to phase $a$}:
{
\begin{align*}
&P_{\text{\upshape{UF}},i}\!\coloneqq\!\frac{1}{3}\norm{P_i\!-\!\mathbbl{1}_3 p_{a,i}}\!=\!\frac{1}{3}\sqrt{(p_{b,i}-p_{a,i})^2\!+\!(p_{c,i}-p_{a,i})^2}, \\
&Q_{\text{\upshape{UF}},i}\!\coloneqq\!\frac{1}{3}\norm{Q_i-\mathbbl{1}_3 q_{a,i}}\!=\!\frac{1}{3}\sqrt{(q_{b,i}-q_{a,i})^2\!+\!(q_{c,i}-q_{a,i})^2}.
\end{align*}
In other words, $P_{\text{\upshape{UF}},i}$ and  $Q_{\text{\upshape{UF}},i}$ capture the unbalance of phase powers relative to phase $a$.}
\end{definition}
Next, we will present and discuss simulation results for both cases.

\subsection{Case~1: Transmission and distribution connected VSCs}
\begin{figure}[t!]
\centering
\includegraphics[width=0.47\textwidth]{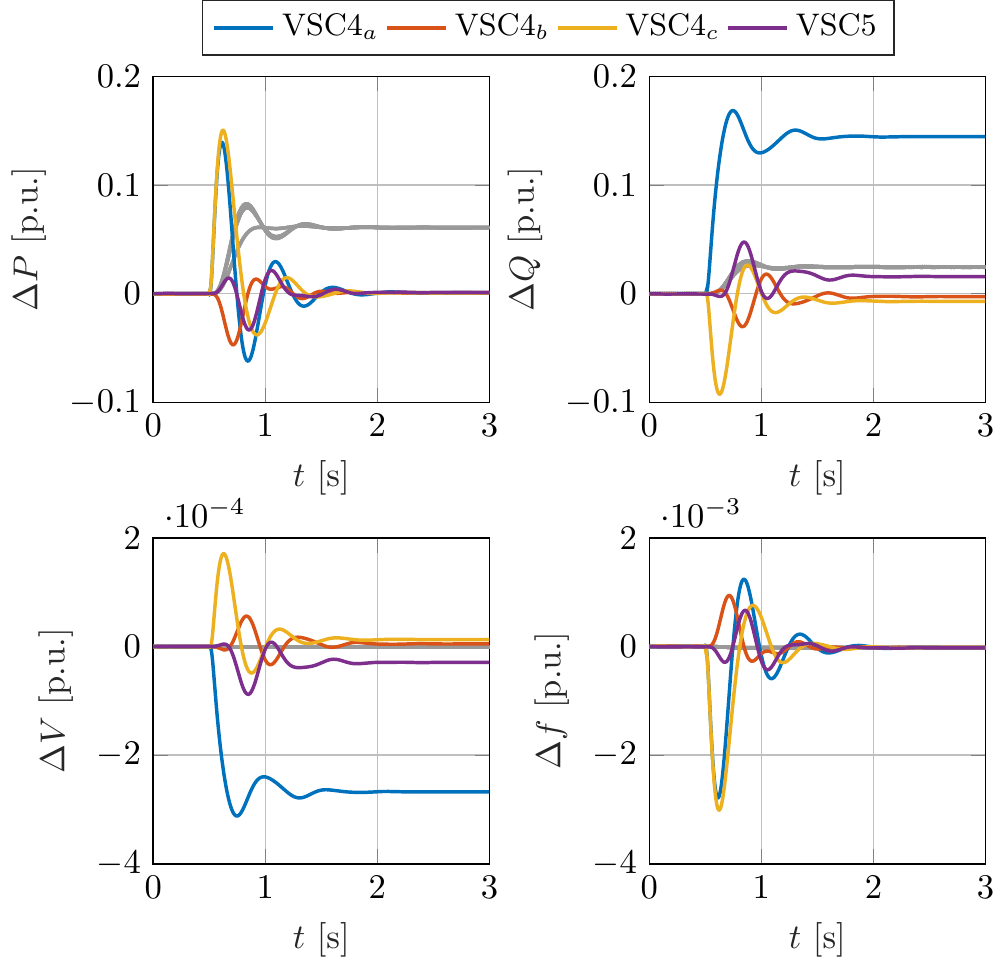}
\caption{Frequency, voltage magnitude, and power when connecting a constant impedance load (approx.  $0.6~\mathrm{p.u.}$ active power) between phases $a$ and $c$ of of bus 680 ($t=0.6~\mathrm{s}$)  with $k_{\text{\upshape{bal}}}=0$. The voltages and reactive power injection of the single-phase VSCs are unbalanced in steady-state, the frequency returns to its nominal value and the active power injection is balanced. Signals of VSCs in the IEEE~9-bus transmission system are shown in grey. All values are in their respective system base.\label{fig9}}
\end{figure}

\begin{figure}[t!]
\centering
\includegraphics[width=0.47\textwidth]{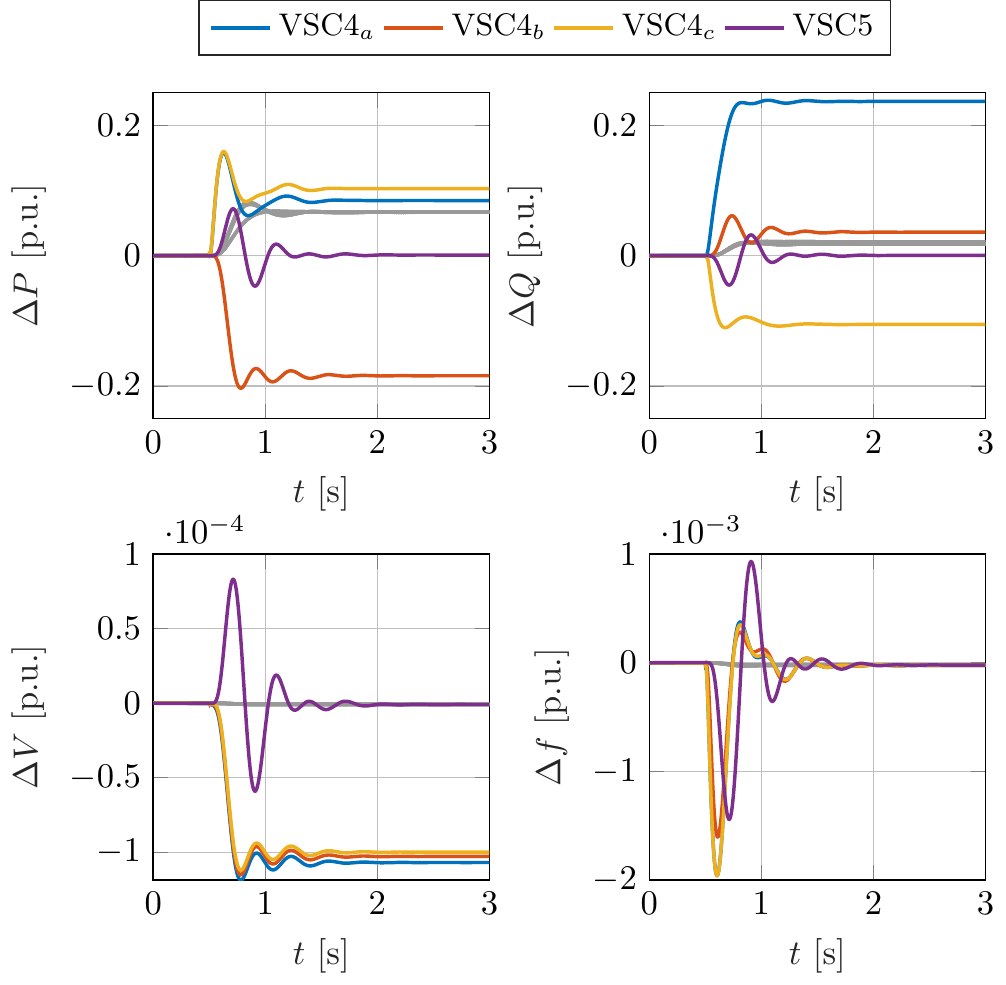}
\caption{Frequency, voltage magnitude, and power when connecting a constant impedance load (approx.  $0.6~\mathrm{p.u.}$ active power) between phases $a$ and $c$ of of bus 680 ($t=0.6~\mathrm{s}$)  with $k_{\text{\upshape{bal}}}=30$. The voltage unbalance at bus 680 is significantly decreased at the expense of unbalanced active and reactive power injection by the single-phase VSCs at bus 680. Signals of VSCs in the IEEE~9-bus transmission system are shown in grey. All values are in their respective system base. \label{fig10}}
\end{figure}

This case study is motivated by systems that contain three-phase converters connected to transmission and single-phase converters connected to distribution. Simulation results for connecting a $0.6~\mathrm{p.u.}$ (active power) constant impedance load between phases $a$ and $c$ of bus 680 at $t=0.6~\mathrm{s}$ are shown in Fig.~\ref{fig9} ($k_{\text{\upshape{bal}}}=0$) and Fig.~\ref{fig10} ($k_{\text{\upshape{bal}}}=30$). Both Fig.~\ref{fig9} and Fig.~\ref{fig10} show synchronization in the network frequency in the steady-state. For $k_{\text{\upshape{bal}}}=0$, Fig.~\ref{fig9} shows that the voltages and reactive power injection of the single-phase VSCs (i.e., VSC~4a, VSC~4b, and VSC~4c) are unbalanced in steady-state. Nonetheless, the frequency returns to its nominal value and the active power injection is balanced. In contrast, using $k_{\text{\upshape{bal}}}=30$, the unbalance of the active and reactive power injection is increased, but the voltage unbalance at bus 680 is significantly decreased (see Fig.~\ref{fig10}). 


\begin{figure}[t!]
\centering
\includegraphics[width=0.47\textwidth]{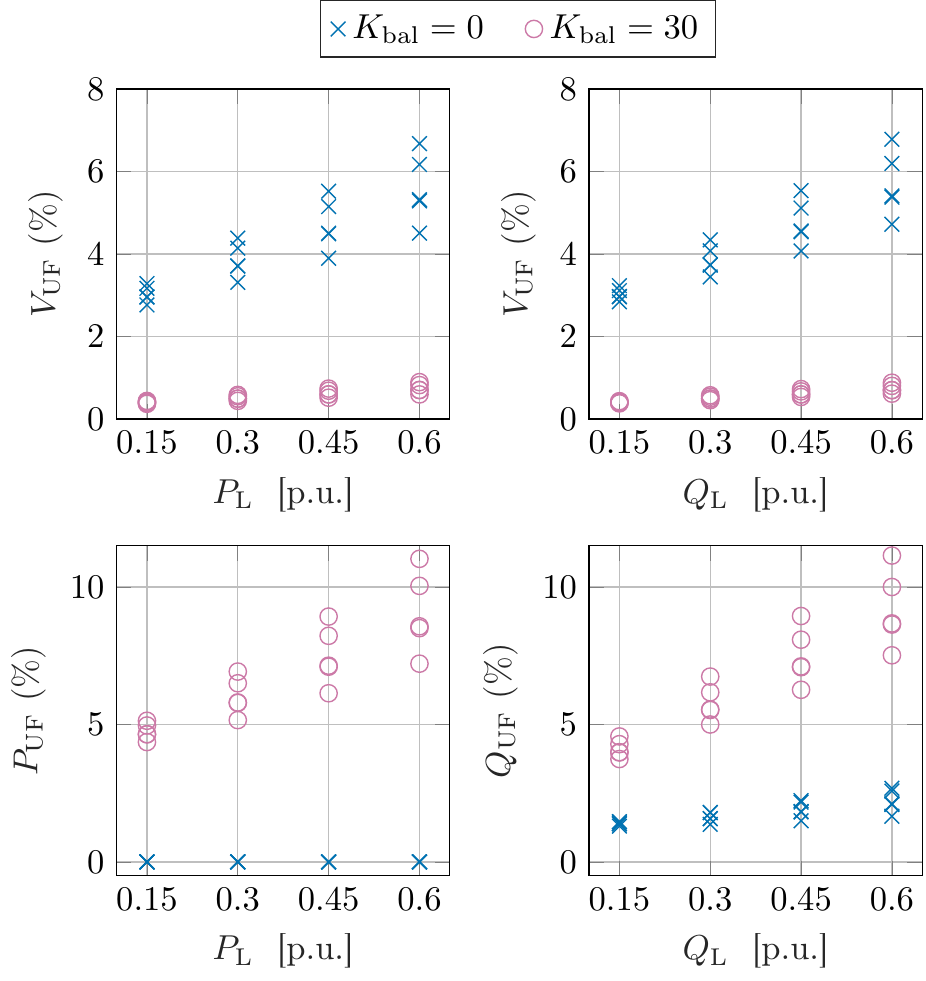}
\caption{Case~1: Unbalance factors for voltage, active power, and reactive power at bus 680 for $k_{\text{\upshape{bal}}}=0$ and $k_{\text{\upshape{bal}}}=30$ and (individually) perturbing the active power and reactive power of $\mathrm{Load~1}$ through $\mathrm{Load~5}$.\label{fig8}}
\end{figure}

\begin{figure}[t!]
\centering
\includegraphics[width=0.47\textwidth]{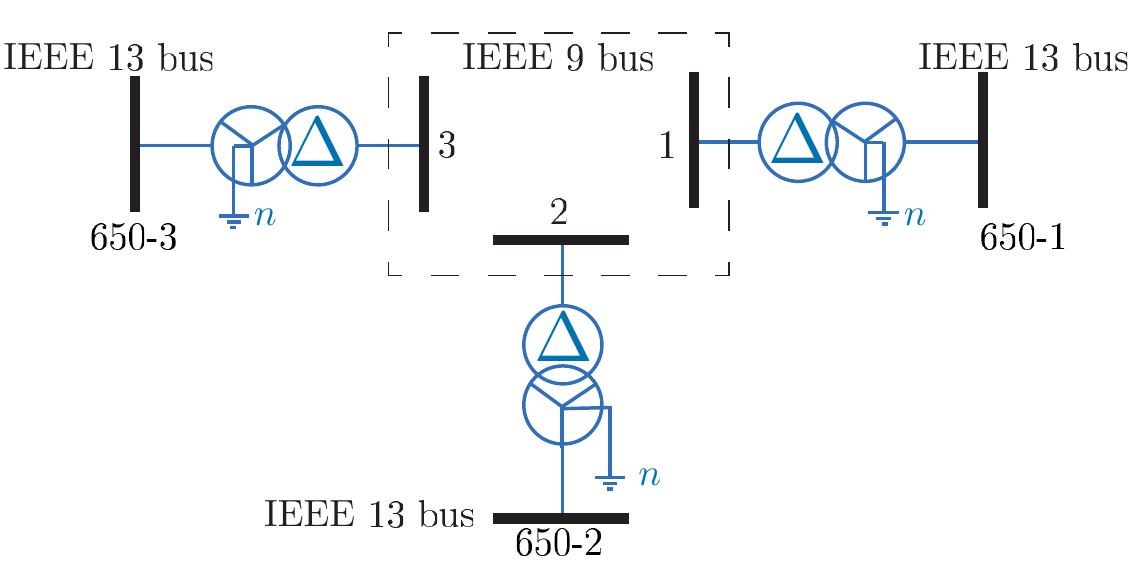}
\caption{Case~2: Three IEEE~13-bus distribution feeders connected to bus 1, bus 2, and bus 3 of the IEEE~9-bus system.}
\label{fig72}
\end{figure}

Next, we investigate the steady-state response to unbalanced load conditions. To this end, the active and reactive power of $\mathrm{Load~1}$ trough $\mathrm{Load~5}$ (impedance loads between phases $a$ and $c$) are varied from $0.15~\mathrm{p.u.}$ to $0.6~\mathrm{p.u.}$ and the voltage and power unbalance factors $V_\text{UF}$, $P_\text{UF}$, and $Q_\text{UF}$ in steady state are shown in Fig.~\ref{fig8}.
It can be seen that $K_{\text{\upshape{bal}}}=0$ results in significant voltage unbalance factors for both unbalanced active power and unbalanced reactive power loads. For instance, connecting a $0.15~\mathrm{p.u.}$ load between phases $a$, and $c$, results in $V_\text{UF}=3\%$. This already exceeds typical recommendations such as maintaining $<1\%$ voltage unbalance to ensure efficient operation of electric motors (see, e.g., NEMA MG-1-1998) or normal operating ranges defined by utilities (e.g., $2.5\%$ in \cite{PGE}). Using $k_{\text{\upshape{bal}}}=30$ for VSC~4, the phase voltages are balanced and the voltage unbalance factor $V_\text{UF}$ is kept below $1\%$ at the expense of unbalance in the active and reactive power injection. In other words, using balancing feedback increases $P_\text{UF}$ and/or $Q_\text{UF}$ of the VSC. For instance, using $K_{\text{\upshape{bal}}}=0$, $P_\text{UF}$ is at $0\%$ for all $P_{\text{\upshape{L}}}$, while $k_{\text{\upshape{bal}}}=30$ increases $P_\text{UF}$ to approximately $5\%$. $Q_\text{UF}$ is non-zero for $k_{\text{\upshape{bal}}}=0$ due to the $Q-V$ droop response of the VSCs and increases for $K_{\text{\upshape{bal}}}=30$. Finally, Fig.~\ref{fig8} highlights the fact that the response of droop control with balancing feedback is approximately linear, i.e., $P_\text{\upshape{UF}}$ and $Q_\text{UF}$ increase approximately linearly as the load and voltage unbalance increases.

\subsection{Case~2: Distribution connected single-phase VSCs}
\begin{figure}[t!]
\centering
\includegraphics[width=0.47\textwidth]{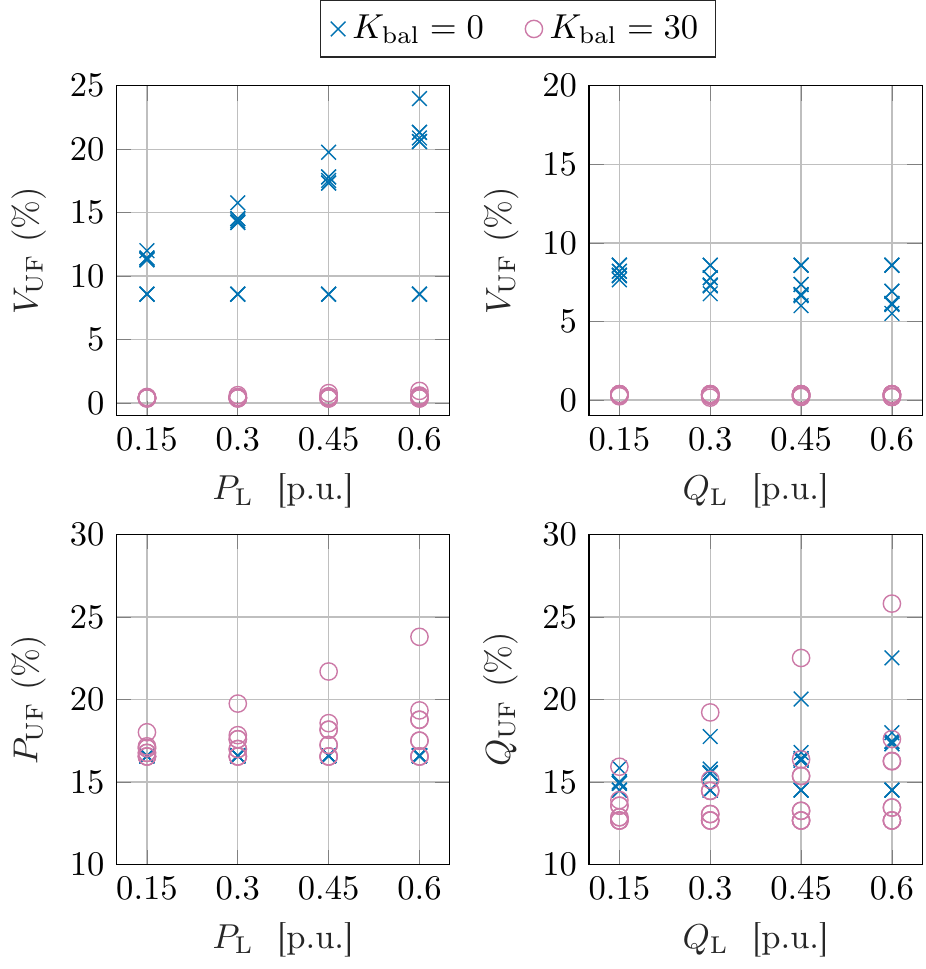}
\caption{Case~2: Unbalance factors for voltage, active power, and reactive power at bus 680-1 for for (individual) perturbations of the active power and reactive power of $\mathrm{Load~1}$ through $\mathrm{Load~5}$ and $k_{\text{\upshape{bal}}} \in \{0,30\}$.\label{fig11}}
\end{figure}

This case represents an extreme scenario in which all generation is interfaced to the distribution system through single-phase VSCs. To study the performance of the generalized three-phase droop control \eqref{eq:droop_3phi}, we again vary both active and reactive power of $\mathrm{Load~1}$ to $\mathrm{Load~5}$ (constant impedance loads between phases $a$ and $c$) from $0.15~\mathrm{p.u.}$ to $0.6 ~\mathrm{p.u.}$ and calculate $V_\text{UF}$, $P_\text{UF}$, and $Q_\text{UF}$ in steady state at the buses 680-1, 680-2, and 680-3. {Fig.~\ref{fig11}} shows the results for the bus 680-1 and $k_{\text{\upshape{bal}}}=0 $ and $k_{\text{\upshape{bal}}}=30$.
For $k_{\text{\upshape{bal}}} =0$, severe voltage unbalances are observed. For instance, for a $0.15 \mathrm{p.u.}$ load between phases $a$, $c$, we obtain $V_\text{UF}=10\%$. In contrast,  using $k_{\text{\upshape{bal}}}=30$), the voltage unbalance factor $V_\text{UF}$ is close to $0\%$. At the same time, the droop response of the VSCs already results in significant power unbalances even in the case $k_{\text{\upshape{bal}}} =0$ (e.g.,  $P_\text{UF}$ is close to $17\%$) due to the absence of any generation with voltage balancing capabilities. Interestingly, using $k_{\text{\upshape{bal}}} =30$ does not significantly change the reactive power unbalance $Q_\text{UF}$ and  $P_\text{UF}$ increases significantly less than in Case~1. For example, connecting a $0.15~ \mathrm{p.u.}$ active load increases the power unbalance $P_\text{UF}$ by approximately $2\%$.

\begin{figure}[t!]
\centering
\includegraphics[width=0.47\textwidth]{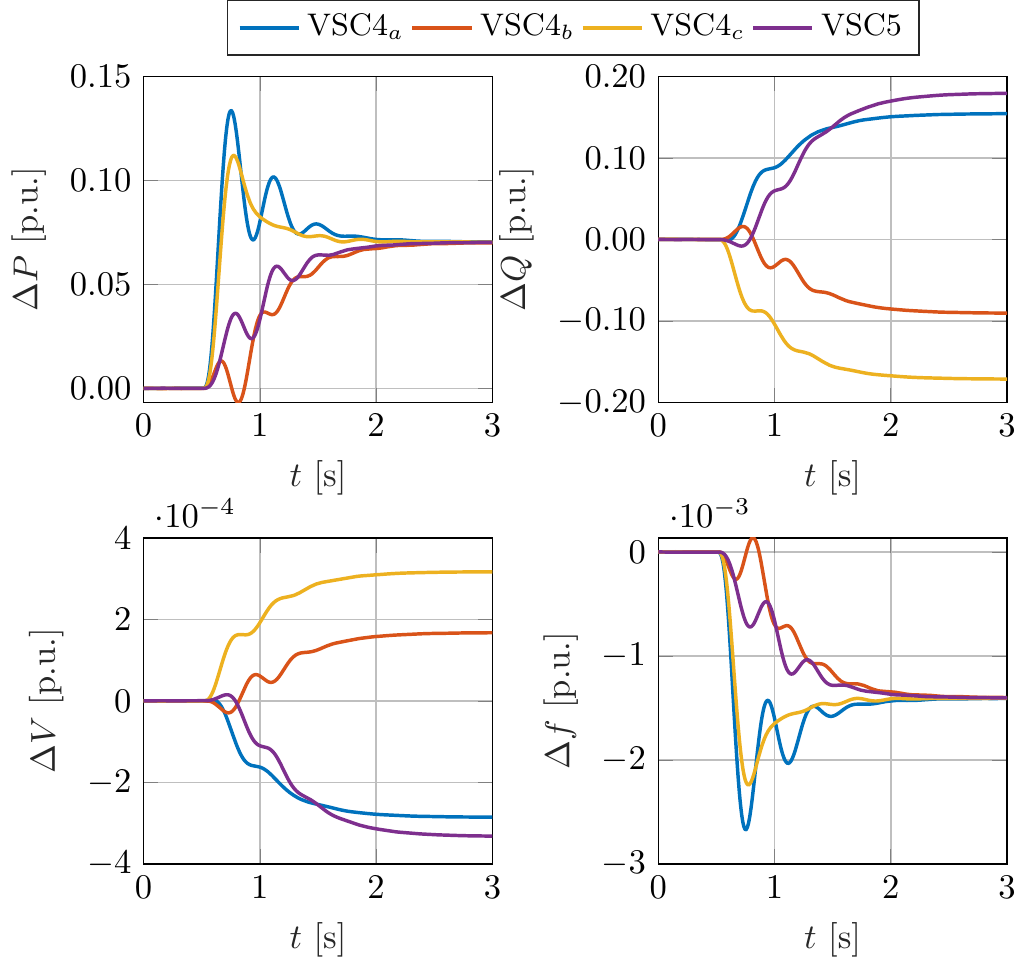}
\caption{Frequency, voltage magnitude, and power when connecting a constant impedance load (approx.  $0.6~\mathrm{p.u.}$ active power) between phases $a$ and $c$ of of bus 680-1 ($t=0.6~\mathrm{s}$)  with $k_{\text{\upshape{bal}}}=0$. The voltages and reactive power injection of the single-phase VSCs are unbalanced in steady-state, the frequency returns to its nominal value and the active power injection is balanced.\label{fig12}}
\end{figure}

\begin{figure}[t!]
\centering
\includegraphics[width=0.47\textwidth]{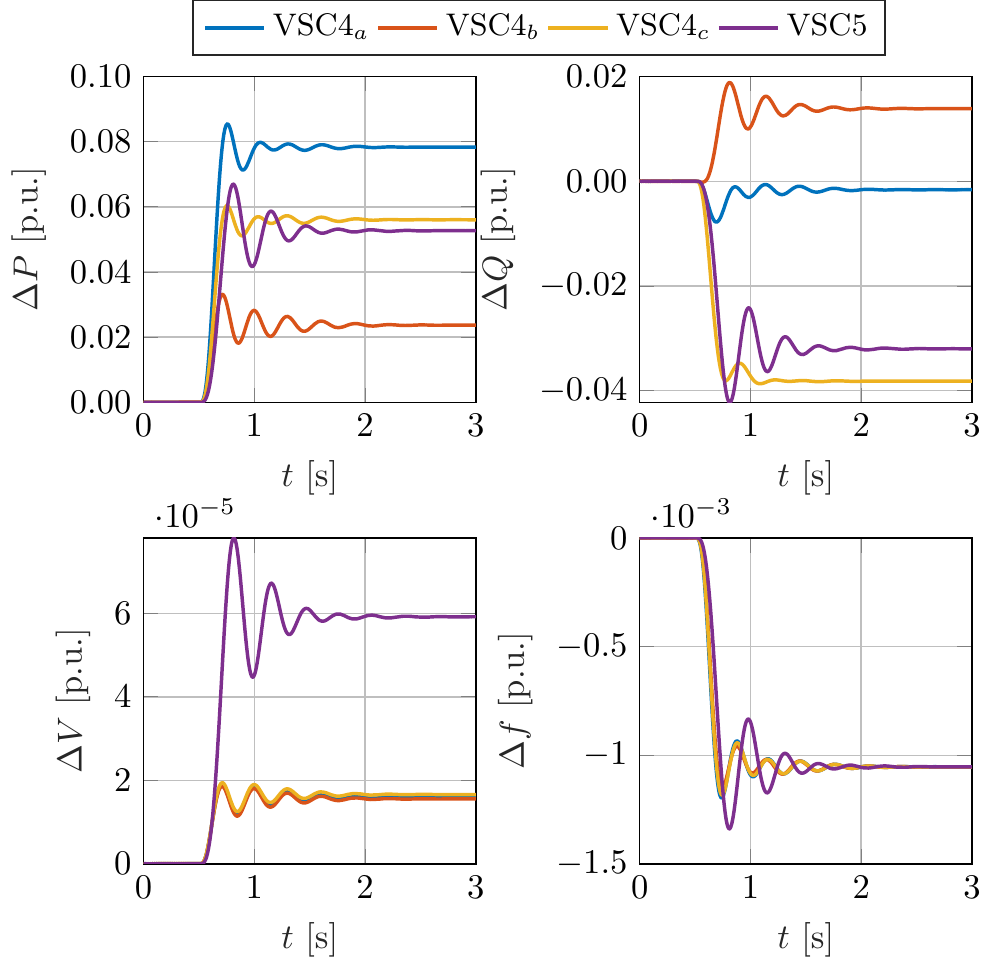}
\caption{Frequency, voltage magnitude, and power when connecting a constant impedance load (approx.  $0.6~\mathrm{p.u.}$ active power) between phases $a$ and $c$ of of bus 680-1 ($t=0.6~\mathrm{s}$)  with $k_{\text{\upshape{bal}}}=30$. The voltage unbalance at bus 680-1 is significantly decreased at the expense of unbalanced active and reactive power injection by the single-phase VSCs at bus 680-1. \label{fig13}}
\end{figure}

The response to connecting a constant impedance load (active power $0.6~\mathrm{p.u.}$) between phases $a$ and $c$ of bus 680-1 is shown in Fig.~\ref{fig12} ($K_{\text{\upshape{bal}}}=0$) and Fig.~\ref{fig13} ($K_{\text{\upshape{bal}}}=30$). The results in line with the observations in Case~1. For $k_{\text{\upshape{bal}}} =0$, the voltages and reactive power injections of the single-phases sources are unbalanced. However frequencies synchronize in the entire network and the active power generation at bus 680-1 is balanced (i.e., $P_\text{UF}=0\%$).
With the balancing feedback, the voltages becomes balanced in the bus 680 in steady-state. However, this results in unbalanced power generations between $\mathrm{VSC}~4a$. $\mathrm{VSC}~4b$, and $\mathrm{VSC}~4c$ and as a result both $P_\text{UF}$ and $Q_\text{UF}$ are increased compared to the unbalanced case.

\subsection{Comparison of voltage unbalance factors}
\begin{figure}[t!]
\centering
\includegraphics[width=0.47\textwidth]{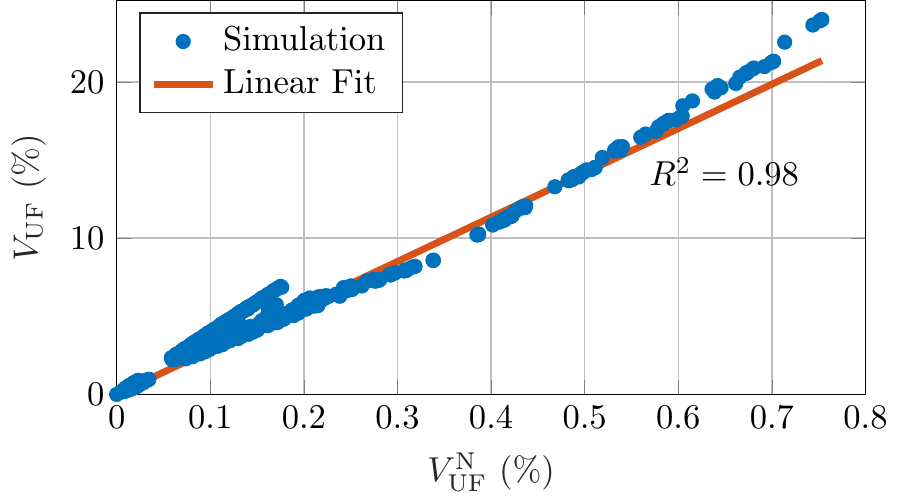}
\caption{Correlation between the voltage unbalance factors $V_\text{UF}$ and $V^\text{N}_{\text{UF}}$ across all simulation results.\label{fig14}}
\end{figure}

Finally, we note that, by design of the balancing feedback, the generalized three-phase droop control \eqref{eq:droop_3phi} feeds back the pairwise difference of voltage phase angle and voltage magnitude differences between phases, i.e., $\mathsf{S}^\mathsf{T} V_{\delta,i}$. To measure the magnitude and angle difference to phase a, we define 
\begin{align} \label{eq:1000}
V^\text{\upshape{N}}_{\text{\upshape{UF}},i}=\frac{1}{3}\left\| \begin{bmatrix} v_i - \mathbbl{1}_3 v_{a,i} \\ \theta_i - \mathbbl{1}_3 \theta_{a,i} \end{bmatrix} \right\|.
\end{align}
While this metric is different from the standard voltage unbalance factor $V_\text{UF}$, we observe that $V_\text{UF}$ and $V^\text{N}_{\text{UF}}$ are highly correlated. Specifically, for all simulation results discussed in this section, we calculated $V_{\text{\upshape{UF}}}$ and $V^\text{\upshape{N}}_{\text{\upshape{UF}}}$ at bus 680 (respectively 680-1, 680-2, and 680-3). The results are shown in Fig.~\ref{fig14}. It can be seen that $V_{\text{\upshape{UF}}}$ and $V^\text{\upshape{N}}_{\text{\upshape{UF}}}$ are strongly correlated for all of our simulations. This observation justifies the design of the balancing feedback as well as the definitions of $P_\text{UF}$ and $Q_\text{UF}$.

\section{Summary and outlook} \label{sec:conclusion}
{This paper studied dynamic stability of converter-dominated power systems that contain a mix of three-phase and single-phase grid-forming converters across transmission and distribution. To this end, we first developed a quasi-steady-state network model that captures the main salient features of three-phase and single-phase networks interconnected through standard transformer interconnections and has suitable complexity for analytic stability studies. Next, we characterized the class of networks for which network reduction techniques (i.e., Kron reduction) can be generalized to unbalanced three-wire/three-phase networks with common transformer interconnections. 

Our main contribution are sufficient conditions for frequency and voltage stability as well as phase-balancing of three-phase and single-phase power converters deployed across transmission and distribution systems. In particular, our analytical stability conditions characterize the network topologies for which (i) single-phase converters phase-balance through synchronization with three-phase converters, and (ii) for which autonomous phase-balancing of single-phase converters occurs. Moreover, we proposed a phase-balancing feedback that ensures stability of the system under mild conditions.} Finally, a detailed case study that combines a three-phase transmission system (IEEE 9-bus) with a distribution system (IEEE 13-bus) is used to illustrate the analytical results and study the proposed balancing feedback. Specifically, we considered two scenarios: (i) a mix of transmission-connected three-phase converters and distribution-connected single-phase converters, (ii) all converters are distribution-connected and single-phase. {Investigating phase-balancing between single-phase converters that are not located at the same bus and extending the results to distribution lines with non-negligible resistance are seen as interesting topics for future work.}

\bibliographystyle{IEEEtran}
\bibliography{IEEEabrv,ref.bib}



\end{document}